\documentclass[twocolumn]{aastex631}

\usepackage{multirow}
\usepackage{amsmath}
\usepackage[utf8]{inputenc}
\usepackage[T1]{fontenc}
\usepackage{soul}
\shorttitle{Outflow in HE~0040$-$1105}
\shortauthors{Singha et al.}

\graphicspath{{./}{figures/}}

\begin{document}

\title{The Close AGN Reference Survey (CARS)\\ 
An interplay between radio jets and AGN radiation in the radio-quiet AGN HE~0040$-$1105}

\author[0000-0001-5687-1516]{M.~Singha}

\affiliation{Astrophysics Science Division, NASA Goddard Space Flight Center, Greenbelt, MD 20771, USA}

\affiliation{Department of Physics, The Catholic University of America, Washington, DC 20064, USA}

\affiliation{Center for Research and Exploration in Space Science and Technology, NASA Goddard Space Flight Center, Greenbelt, MD 20771, USA}

\affiliation{Department of Physics \& Astronomy, University of Manitoba, 30A Sifton Road, Winnipeg, MB R3T 2N2, Canada}

\author[0000-0001-9428-6238]{N.~Winkel}
\affiliation{Max-Planck-Institut f\"{u}r Astronomie, K\"onigstuhl 17, D-69117 Heidelberg, Germany}

\author[0000-0003-3295-6595]{S.~Vaddi}
\affiliation{Arecibo Observatory, NAIC, HC3 Box 53995, Arecibo, Puerto Rico, PR 00612, USA}

\author[0000-0001-5654-0266]{M.~Perez Torres}
\affiliation{Instituto de Astrof\'isica de Andaluc\'ia (IAA-CSIC), Glorieta de la Astronom\'ia s/n, E-18008 Granada, Spain}
\affiliation{Facultad de Ciencias, Universidad de Zaragoza, Pedro Cerbuna 12, E-50009 Zaragoza, Spain}
\affiliation{School of Sciences, European University Cyprus, Diogenes Street, Engomi, 1516 Nicosia, Cyprus}

\author[0000-0003-2754-9258]{M.~Gaspari}
\affiliation{Department of Astrophysical Sciences, Princeton University, 4 Ivy Lane, Princeton, NJ 08544-1001, USA}

\author[0000-0002-2260-3043]{I.~Smirnova-Pinchukova}
\affiliation{Max-Planck-Institut für Astronomie, K\"onigstuhl 17, D-69117 Heidelberg, Germany}

\author[0000-0001-6421-054X]{C.~P.~O'Dea}
\affiliation{Department of Physics \& Astronomy, University of Manitoba, 30A Sifton Road, Winnipeg, MB R3T 2N2, Canada}

\author[0000-0003-2658-7893]{F.~Combes}
\affiliation{LERMA, Observatoire de Paris, PSL Research Univ., College de France, CNRS, Sorbonne Univ., UPMC, Paris, France}

\author{O.~Omoruyi}
\affiliation{Center for Astrophysics $|$ Harvard \& Smithsonian, 60 Garden St., Cambridge, MA 02138, USA}

\author[0000-0002-8310-2218]{T.~Rose}
\affiliation{Department of Physics and Astronomy, University of Waterloo, Waterloo, ON N2L 3G1, Canada}
\affiliation{Waterloo Centre for Astrophysics, University of Waterloo, Waterloo, ON N2L 3G1, Canada}

\author[0000-0002-7960-5808]{R.~McElroy}
\affiliation{School of Mathematics and Physics, University of Queensland, St Lucia, QLD 4072, Australia}

\author[0000-0003-2901-6842]{B.~Husemann}
\affiliation{Max-Planck-Institut für Astronomie, K\"onigstuhl 17, D-69117 Heidelberg, Germany}

\author[0000-0003-4932-9379]{T.~A.~Davis}
\affiliation{Cardiff Hub for Astrophysics Research \&\ Technology, School of Physics \&\ Astronomy, Cardiff University, The Parade, Cardiff, CF24 3AA, UK}

\author[0000-0002-4735-8224 ]{S.~A.~Baum}
\affiliation{Department of Physics \& Astronomy, University of Manitoba, 30A Sifton Road, Winnipeg, MB R3T 2N2, Canada}

\author[0000-0002-2958-0593]{C.~Lawlor-Forsyth}
\affiliation{Department of Physics and Astronomy, University of Waterloo, Waterloo, ON N2L 3G1, Canada}
\affiliation{Waterloo Centre for Astrophysics, University of Waterloo, Waterloo, ON N2L 3G1, Canada}

\author[0000-0002-3289-8914]{J.~Neumann}
\affiliation{Max-Planck-Institut f\"{u}r Astronomie, K\"onigstuhl 17, D-69117 Heidelberg, Germany}

\author[0000-0002-5445-5401]{G.~R.~Tremblay}
\affiliation{Center for Astrophysics $|$ Harvard \& Smithsonian, 60 Garden St., Cambridge, MA 02138, USA}

\correspondingauthor{M.~Singha}
\email{singham@myumanitoba.ca}

\begin{abstract}
\textcolor{black}{We present a case study of HE~0040$-$1105, an unobscured radio-quiet AGN at a high accretion rate $\lambda_{Edd} = 0.19\pm0.04$. This particular AGN hosts an ionized gas outflow with the largest spatial offset from its nucleus compared to all other AGNs in the Close AGN Reference Survey (CARS). By combining multi-wavelength observations from VLT/MUSE, HST/WFC3, VLA, and EVN we probe the ionization conditions, gas kinematics, and radio emission from host galaxy scales to the central few pc. We detect four kinematically distinct components, one of which is a spatially unresolved AGN-driven outflow located within the central $500~\text{pc}$, where it locally dominates the ISM conditions. Its velocity is too low to escape the host galaxy's gravitational potential, and maybe re-accreted onto the central black hole via chaotic cold accretion. We detect compact radio emission in HE~0040$-$1105 within the region covered by the outflow, varying on $\sim 20$ yr timescale.
We show that neither AGN coronal emission nor star formation processes wholly explain the radio morphology/spectrum. The spatial alignment between the outflowing ionized gas and the radio continuum emission on $100~\text{pc}$ scales is consistent with a weak jet morphology rather than diffuse radio emission produced by AGN winds. 
$> 90\%$ of the outflowing ionized gas emission originates from the central $100~\text{pc}$, within which the ionizing luminosity of the outflow is comparable to the mechanical power of the radio jet. Although radio jets might primarily drive the outflow in HE~0040$-$1105, radiation pressure from the AGN may contribute in this process.}
\end{abstract}

\section{Introduction}

Active galactic nuclei (AGN) are capable of expelling ionized gas from the centers of galaxies with velocities that reach several hundred to tens of thousands of $\text{km s}^{-1}$ through a process known as AGN-driven outflows. Resolving the underlying physical processes behind these outflows is crucial for understanding galaxy evolution, as they are required to explain how AGN feed back the energy from accreted materials to their host galaxies \citep{Fabian2012,Gaspari:2020,Hardcastle2020}, and produce tight positive correlations between the supermassive black hole (SMBH) mass and host galaxy properties \citep[e.g.,][]{Silk1998,King2005,Kormendy:2013,Gaspari:2019}. Important questions that need to be resolved include how outflows are launched from the parsec-scale central engine, how they propagate through the host galaxy to reach kpc-scales, and how their energy thereby couples to the surrounding interstellar medium (ISM).

Outflows in the warm, ionized gas phase can be traced by the broad blue-shifted wing components in the forbidden [\ion{O}{3}] emission line doublet and are often detected in AGN \citep{Bischettii2017,cresci2015,Singha2021}. The statistical power of 1D optical spectroscopic data sets of large sky surveys has helped to quantify their abundance and energetics. The spatial scales are particularly important to consider, as recent studies have shown that in many cases the luminous [\ion{O}{3}]-emitting outflow is more compact than expected \citep[e.g.,][]{Husemann2016,Singha2021a, Winkel:2022b}. Spatially resolving the ionized gas outflow is therefore crucial to understand the local impact on the host galaxy. Moreover, the nuclear outflow launching mechanism and the connection to the AGN phase remain elusive.

One scenario has been suggested, where ionized gas outflows are launched from pc-scales by the radiation pressure from the AGN accretion disk. \citet{Woo2016} and \citet{Rakshit2018} found that the incidence of ionized gas outflows is predominantly correlated with the AGN bolometric luminosity, which suggests that radiation pressure is the main driver of outflows. A recent work by \citet{Singha2021} which focused on weakly accreting (Eddington ratio $\lambda_{\text{Edd}} < 0.01$), low-excitation radio AGN (LERGs) also hinted towards radiation pressure from the AGN accretion disk driving sub-kpc scale ionized gas outflows. The foundation for an AGN-wind scenario comes from numerous theoretical studies \citep{Elvis2000,King2005,Proga2007,Faucher-Giuere12}, followed by observations \citep{Zakamska2014,Baron2019DiscoveringLocation,Sun2019,U2022}. They describe the momentum transfer from AGN-driven winds as the underlying physical mechanism that allows the ionized gas to expand. Theoretical studies such as \citet{King2003} and \citet{Faucher-Giuere12} mentioned that the high-velocity wind-driven outflows would be energy-conserving on large scales but momentum-conserving near their launch locations.

Over the last three decades, several studies have suggested a possible connection between compact radio jets and multi-phase gas outflows. Studies by \citet{Whittle1992,Tadhunter2003,Holt2008,Morganti2013} suggested that the mechanical energy of compact radio jets could strongly perturb and accelerate the ambient multi-phase gas, creating bi-polar outflows extended on sub-kpc scales. \citet{Mullaney2013,Molyneux2019} reported that the compact radio jets could strongly interact with the ambient gas, resulting in a large line width of [\ion{O}{3}] emitting ionized gas clouds, $\text{FWHM}_{\text{[O~III]}} > 1000~\text{km s}^{-1}$, originating from turbulent outflowing gas. This turbulence is a key element in the recurrent cycle of AGN feedback and feeding, in particular further stimulating the chaotic cold accretion (CCA) rain of multi-phase gas at different scales \citep[e.g.,][]{Gaspari:2018}, thus inducing spectral line broadening.

It is clear that radio jets can shock, accelerate, and entrain the ambient gas in radio-loud AGN \citep{Baum1997,Emonts2005,Laing2014,Mahony2016,Schulz2018}, the exact role of jets in radio-quiet AGN remains unclear.  For a review of jets in radio-quiet AGN and their role in feedback see \citet{Singha2023}.
Unlike radio-loud AGN, radio-quiet AGN, which lack the kpc-scale extended radio jet emission, constitute $\sim 90\%$ of the AGN population. However, radio interferometric observations by \citet{Miller1993,Ulvestad2005,Gallimore2006,Berton2020} showed that compact radio jets are present on sub-kpc scales even in radio-quiet AGN.

\citet{Panessa2019} discussed different mechanisms of radio emission which include star formation and the associated supernovae feedback, black hole coronal emission, AGN winds, and radio jets. \citet{Laor2008} have argued that processes induced by the magnetic field in the AGN corona which causes non-thermal synchrotron emission can create ionized gas outflows. In contrast, \citet{Blundell1998} proposed the observed radio emission in radio-quiet AGN is due to the thermal free-free emission from radiatively driven AGN winds. \citet{Zakamska2014} reported that the shock-accelerated electrons by the AGN winds cause synchrotron emission. \citet{Aalto2017,Husemann2019,Jarvis2019,Girdhar2022} suggested that small-scale radio jets could drive multi-phase gas outflows in radio-quiet AGN. The diversity of these mechanisms demonstrates the complexity of understanding the launching mechanism in radio-quiet AGN. Nevertheless, these results emphasize that small scales close to the nucleus must be probed to distinguish between the different mechanisms. Thus, spatially resolved multi-wavelength observations of AGN-driven outflows are crucial to determine their properties in detail and identify their powering mechanisms \citep[e.g.,][]{Cicone2018}.

The Close AGN Reference Survey\footnote{\url{https://www.cars-survey.org}} \citep[CARS;][]{Husemann2017,McElroy2022}, consists of a sample of 40 unobscured, radio-quiet AGN which represents the majority of the nearby ($z < 0.06$) and luminous ($L_{\text{bol}} < 10^{46}~\text{erg~s}^{-1}$) AGN population well. \citet{Singha2021a} showed that in the majority ($63\%$) of the CARS sample that have ionized gas outflows in [\ion{O}{3}] (blue wing) are spatially unresolved and their flux-weighted centroids are located within $100~\text{pc}$ from the AGN nucleus. Moreover, the authors found that the occurrence of kpc-scale outflows is not correlated with $L_{\text{bol}}$, indicating that the radio emission or outflow timescales affect the outflows. An unambiguous understanding of the connection between radio emission and ionized gas outflows can only be made by spatially resolving both the radio and ionized gas emission on sub-kpc scales.

In this work we carry out a pilot study of HE~0040$-$1105 \citep[\textcolor{black}{RA = 00:42:36.9, DEC = -10:49:22};][]{Husemann2021}, a nearby ($z = 0.04196$), luminous ($L_{\text{bol}} \sim 10^{44}~\text{erg s}^{-1}$) Seyfert~1 galaxy that hosts a radio-quiet, unobscured AGN. Due to its proximity (angular distance $D_{\text{A}} \sim 171~\text{Mpc}$), HE~0040$-$1105 gives us a detailed view of the AGN-host galaxy interaction at a high spatial resolution where $1\arcsec$ corresponds to $828~\text{pc}$ in the galaxy system. The host galaxy is an unbarred bulge-dominated galaxy with $M_{\text{B}} = -19.38~\text{mag}$ and a stellar mass $\log{(M_{*}/\text{M}_{\odot})} = 10.16_{-0.10}^{+0.13}$. Soft X-ray emission has been detected by ROSAT within the $0.1 - 2.4~\text{keV}$ range as well, with $f_{0.1 - 2.4~\text{keV}} = (7.36 \pm 0.81) \times 10^{-12}~\text{erg s}^{-1}~\text{cm}^{-2}$ \citep{Husemann2021}.

HE~0040$-$1105's [\ion{O}{3}] wing component is spatially unresolved at the resolution of VLT/MUSE and the flux weighted centroid of the blue wing [\ion{O}{3}] is $\sim 90~\text{pc}$ offset from the nucleus \citep{Singha2021a}, which represents the largest offset among the CARS AGN with unresolved outflows. By combining the 3D spectroscopic data with radio interferometric observations, we aim to trace the kinematic and spatial components of the outflow and discuss its potential origin.

This paper is organized as follows: In Section~\ref{sec:obs}, we first briefly describe our observations and data reduction. We then investigate the ionized gas properties and radio emission in Section~\ref{sec:analysis}, while in Section~\ref{sec:discussion} we attempt to explain the origin of any observed radio emission mechanisms and their potential connection to the outflows. We will also discuss the ionization mechanism of the emission line gas clouds. Finally in Section~\ref{sec:conclusion} we present our conclusions.

We define the spectral index $\alpha$ as $S_{\nu} \propto \nu^{\alpha}$ where $S_{\nu}$ is the flux density and $\nu$ is the frequency. Throughout this paper we adopt the standard flat $\Lambda$CDM cosmology with $H_{0} = 70~\text{km s}^{-1}~\text{Mpc}^{-1}$, $\Omega_{\text{m}} = 0.3$, and $\Omega_{\Lambda} = 0.7$.

\section{Data}\label{sec:obs}

\begin{deluxetable*}{ccCCCCC}
    \tablecaption{Summary of observational characteristics and depth. VLT/MUSE provides IFU spectroscopy at $1\arcsec$ resolution between $4800-9300~\text{\AA}$, covering a $1\arcmin \times 1\arcmin$ region. VLA has very similar FoV as MUSE but is capable of imaging at $400~\text{mas}$ and $200~\text{mas}$ resolution between $4-8$ and $8-12~\text{GHz}$. EVN resolves the nuclear region of the galaxy at $10~\text{mas}$ resolution.}
    \label{table:ObsRef}
    \tablehead{
    \colhead{Band} & 
    \colhead{Instrument} & 
    \colhead{FoV} & 
    \colhead{Sampling} &
    \colhead{Beam} & 
    \colhead{$t_{\text{exp}}$} & 
    \colhead{$1\sigma$ limit}
    }
    \startdata
    H$\alpha$ & VLT/MUSE & 1\arcmin \times 1\arcmin & 0\farcs2 & 0\farcs66 \times 0\farcs66 & 800~\text{s} & 5 \times 10^{-18}~\text{erg s}^{-1}~\text{cm}^{-2}~\text{arcsec}^{-2} \\
    {[\ion{O}{3}]} & VLT/MUSE & 1\arcmin \times 1\arcmin & 0\farcs2 & 0\farcs74 \times 0\farcs74 & 800~\text{s} & 7.5 \times 10^{-18}~\text{erg s}^{-1}~\text{cm}^{-2}~\text{arcsec}^{-2} \\
     NUV & WFC3/UVIS F336W & 40 \arcsec \times 40 \arcsec & 0\farcs04 & 0\farcs07 & 696~\text{s} & 1.29 \times 10^{-18}~\text{erg}~\text{cm}^{-2}~\text{\AA}^{-1}~\text{e-}^{-1} \\
    C & VLA (A config) & 4.5\arcmin & 0\farcs066 & 0\farcs53 \times 0\farcs27 & 1~\text{hr} & 5.9~\mu\text{Jy beam}^{-1} \\
    X & VLA (A config) & 4.5\arcmin & 0\farcs033 & 0\farcs26 \times 0\farcs17 & 1~\text{hr} & 7.3~\mu\text{Jy beam}^{-1} \\
    L & EVN & 9\farcs9 & 0\farcs0006 & 0\farcs02 \times 0\farcs01 & 7~\text{hr} & 12.05~\mu\text{Jy beam}^{-1} \\
    \enddata
\end{deluxetable*}

We utilize the CARS multi-wavelength data set of HE~0040$-$1105 by combining the 3D spectroscopic optical cubes acquired with VLT/MUSE with radio observations obtained with VLA and EVN. In this section, we briefly describe the observations and the data reduction procedures. We additionally summarize the multi-wavelength observations in Table~\ref{table:ObsRef}.

\subsection{Optical integral-field spectroscopy}

\begin{figure*}
  \includegraphics[width=\textwidth]{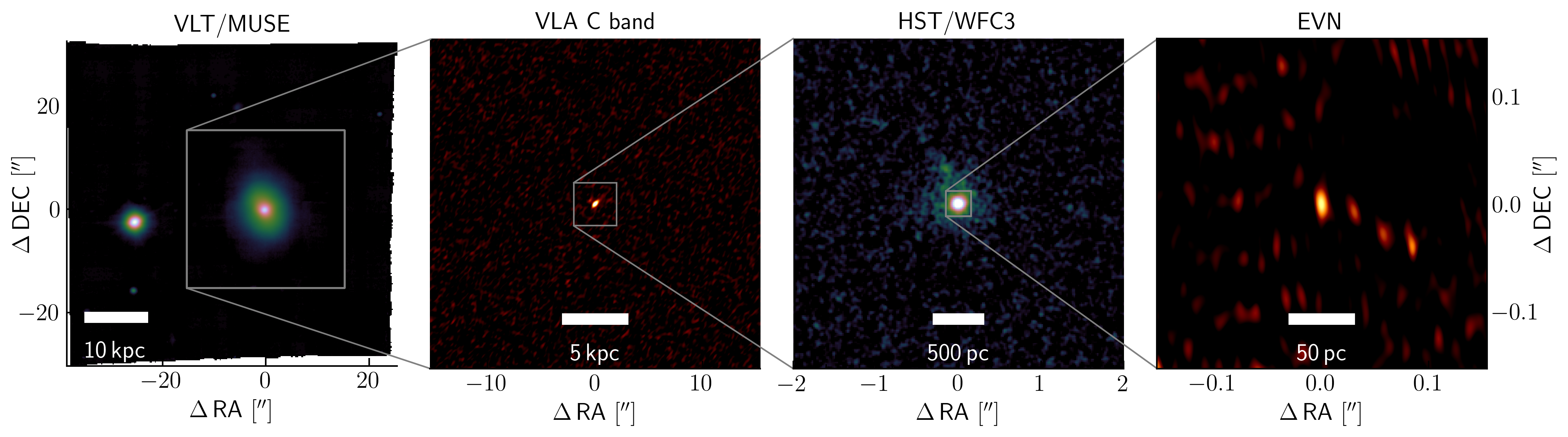}
   \caption{Overview of the data sets and their spatial coverage. From left to right the panels show the field of view of VLT/MUSE, VLA, a zoom-in onto the nucleus in the \textit{HST}/WFC3 UVIS F225W image, and the EVN image. While the MUSE and VLA observations cover a large fraction of the host galaxy, \textit{HST} allows a zoom into the central few $100~\text{pc}$ of the galaxy. The radio interferometry from EVN ultimately resolves the innermost few $10~\text{pc}$ near the AGN. The object $25\arcsec$ east of HE~0040$-$1105 is a foreground star.}
   \label{fig:RGB}
\end{figure*}

The CARS sample is exclusively selected on the $B$-band flux published in the Hamburg-ESO Survey \citep{wisotzki2000}. It is designed such that it covers a representative sample of unobscured luminous AGN at redshifts $0.01 < z < 0.06$. As part of CARS, we have obtained integral field spectroscopic observations of HE~0040$-$1105 with VLT/MUSE in the wide field mode \citep[WFM;][]{Husemann2017,Husemann2021}. The total integration time for HE~0040$-$1105 was $800~\text{s}$ with an effective seeing of $\sim 0\farcs62$. All data were reduced with the standard ESO MUSE pipeline \citep{weilbacher2020} as described in \citet{Husemann2019} and \citet{Husemann2021}. A first visual impression of the reduced data is given in Figure~\ref{fig:RGB}, where we show a broad-band image constructed from the three-dimensional cube. We adopt the model of the stellar continuum emission of the host galaxy retrieved with the stellar population synthesis code \texttt{PyParadise}\footnote{\url{https://github.com/brandherd/PyParadise}}. To analyze the spatially resolved host galaxy stellar and ionized gas component in Section~\ref{subsec:stellar_velocity_field} and Section~\ref{subsec:resolved_host_galaxy_emission}, we employ the stellar continuum subtracted reduced MUSE cube from CARS data release 1 \citep[DR1;][]{Husemann2021}. 

\subsection{NUV imaging}

Near-UV imaging was undertaken using the Wide Field Camera 3 (WFC3) aboard the \textit{Hubble Space Telescope} (\textit{HST}) as part of the Cycle 28 GO program 16173 (PI: G.~R.~Tremblay; Omoruyi et al.~in prep.). We imaged the target using the UVIS detector with the F225W and F336W filters (centered on rest-frame central wavelengths of $\lambda = 2358~\text{\AA}$ and  $3354~\text{\AA}$, respectively) for $696~\text{s}$ (roughly half an orbital visibilty) per filter. A six-point dither pattern was used to both optimally sample and avoid saturation of the PSF from the bright point source associated with the AGN. This PSF has \textit{not} been subtracted from the images shown in this paper, although careful PSF subtraction will be presented in a forthcoming paper by Omoruyi et al. We used the UVIS 1k subarray mode to improve detector readout time and therefore maximize integration time across the dither pattern. The final FoV of each image is therefore $40\arcsec \times 40\arcsec$. The data shown here were reduced using the standard \textit{HST}/WFC3 recalibration pipeline \citep{sahu21} including CTE correction. The \texttt{AstroDrizzle} \citep{Gonzaga2012} algorithm was used to drizzle the final image to a pixel scale of $0.04\arcsec~\text{pixel}^{-1}$ using a pixel droplet fraction of 0.8.

\subsection{Radio interferometric observations}

\subsubsection{Very Large Array imaging}

We acquired radio interferometric observations of HE~0040$-$1105 with the Very Large Array (VLA) on January 7, 2017, and November 22, 2016, in the C- ($4-8~\text{GHz}$) and X- ($8-12~\text{GHz}$) bands, respectively (project: 16B-084, PI: M.~P{\'e}rez Torres). The array was set to the A configuration with a maximum baseline of $36.2~\text{km}$. In both bands, the on-source integration time was $608~\text{s}$. We used 3C~48 to set the flux density scale and J0050$-$0929 as the phase calibrator. We reduced the data with the Common Astronomy Software Applications \citep[CASA, version 6.2.1.7;][]{McMullin2007} and the CASA VLA pipelines before reconstructing the image using the \texttt{tclean} routine with Briggs weighting. The resulting beam sizes along the major axes are $0\farcs5$ (C band) and $0\farcs3$ (X band), which both have an elongation of $0\farcs5$ with respect to the minor axes. The JVLA observations provide sub-arcsec resolution of the radio structures close to the AGN nucleus as shown in Figure~\ref{fig:MUSE_VLA}.

\subsubsection{European VLBI Network imaging}

Very long baseline interferometric observations of HE~0040$-$1105's continuum emission were retrieved with the European VLBI Network (EVN) at $18~\text{cm}$ (project code: EP119, PI: M.~P{\'e}rez Torres) on March 12, 2020, where sixteen stations were used to acquire the observations (Jb, Wb, Ef, O8, Tr, Hh, Sv, Zc, Ir, Sr, Ro, Cm, Da, Kn, Pi, De). The total exposure time was $7~\text{hr}$ with a data recording rate of $1024~\text{Mbps}$ ($8 \times 16~\text{MHz}$ sub-bands, full polarization, two-bit sampling). While J0050$-$0929 was used as a fringe finder, both J0039$-$0942 and J0050$-$0929 were used as phase calibrators. The observations were carried out in the phase-referencing mode where the telescopes were pointed to the target and the phase calibrator in repeated $5~\text{min}$ cycles. During each cycle, $4~\text{min}$ was spent on the target.

The EVN data set was calibrated in the Astronomical Image Processing System (AIPS), a software package developed by the National Radio Astronomy Observatory following the standard EVN data reduction guide \footnote{\url{https://www.evlbi.org/evn-data-reduction-guide}}. The calibration tables that contain parallactic angle a-priori gain corrections were transferred from the EVN pipeline. Additionally, the flag and bandpass tables were also transferred to the data set. The ionospheric dispersive delays were corrected using the VLBATECR task. We performed fringe fitting and bandpass calibration using the fringe finder. Delays and rates were corrected by fringe fitting the data from the phase reference calibrators. The final calibrated target data was exported to DIFMAP for imaging, and the resulting final image was created using natural weighting. The root mean square noise (rms) of the cleaned image is $12~\mu\text{Jy beam}^{-1}$ and the peak intensity is $83.7~\mu\text{Jy beam}^{-1}$.

\begin{figure}
    \hfill\includegraphics[width=\columnwidth]{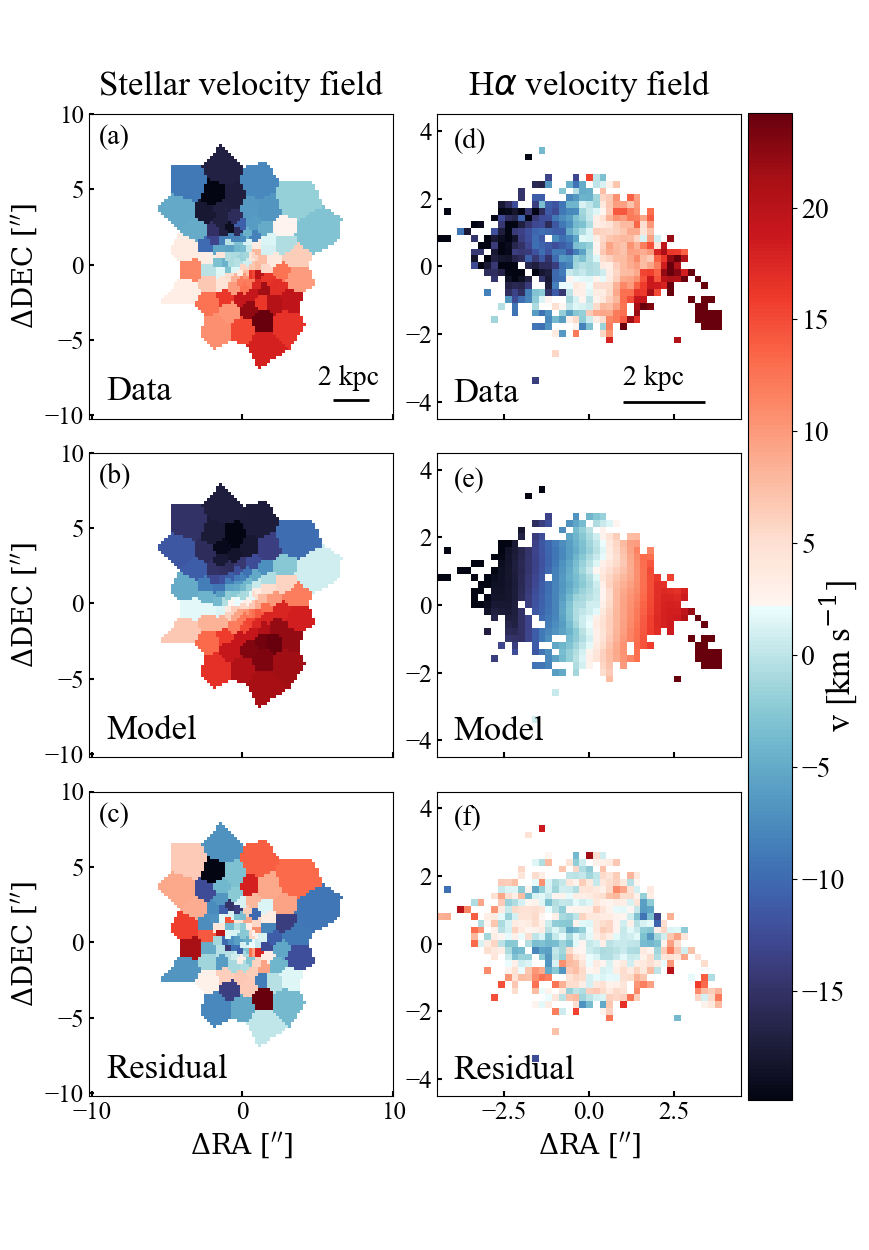}\hspace*{\fill}
    \caption{Kinematic modelling of HE~0040$-$1105's stellar (left columns) and H$\alpha$ ionized gas (right columns) velocity field. From top to bottom the panels show the velocities measured from the MUSE observations (see Section~\ref{subsec:stellar_velocity_field} and Section~\ref{subsubsec:EELR}), the tilted-ring model fitted to it, and the map with the residual velocities normalized by the uncertainty. In both cases, the thin rotating disc provides a good description of the velocity field, although the rotation axis between both host galaxy components is tilted by $53\arcdeg$.}
    \label{fig:stellar_vel_model}
\end{figure}

\section{Analysis and Results}\label{sec:analysis}

We perform a detailed multi-wavelength analysis to investigate the spatial and kinematic properties of the AGN-driven ionized gas outflow, its origin, and its potential effects on the host galaxy. In Section~\ref{subsec:stellar_velocity_field} we analyze the stellar kinematics and measure the host galaxy's systemic redshift. This will allow us to derive the kinematics and energetics of the ionized gas outflow in Section~\ref{subsubsec:outflow_energetics}. In Section~\ref{subsec:resolved_host_galaxy_emission} we then focus on different features that are present in the spatially resolved extended emission line region (EELR) of HE~0040$-$1105. Finally, we analyze the radio interferometric observations in Section~\ref{subsec:radio_emission} to examine whether the processes that drive the ionized gas outflow are related to the resolved radio structures.

\subsection{The host galaxy stellar component}\label{subsec:stellar_velocity_field}

To understand the dynamics of the ionized gas, we first need to constrain the kinematics of HE~0040$-$1105's stellar component. The stellar continuum model from CARS DR1 \citep{Husemann2021,McElroy2022} allows us to map the stellar velocity field that is shown in the upper left-hand panel of Figure~\ref{fig:stellar_vel_model}. We model the 2D velocity field with the tilted ring model as described in \citet{Winkel:2022a}. The code is based on the \texttt{KinMS} package \citep{Davis:2013} which was originally designed to simulate the atomic and molecular gas distributions of galaxies. We assume that the stellar emission originates from a rotating axisymmetric thin disk where the 2D line-of-sight velocity is described by 
\begin{equation}
    v_{\text{r}} = v_{\text{sys}} + v \sin{(i)}\cos{(\phi + \text{PA})},
\end{equation}
where $v_{\text{sys}}$ is the systemic velocity, PA is the position angle of the rotation axis, and $\phi$ is the azimuthal angle measured around the AGN position in the observed plane. We then use the \texttt{GAStimator} algorithm\footnote{\url{https://github.com/TimothyADavis/GAStimator}} to maximize the log-likelihood of the line-of-sight velocity distribution and radially evaluate the model in concentric aperture rings. For this step we only adopt the bins where the uncertainty in the radial velocity is smaller than $10~\text{km s}^{-1}$. Assuming a smooth rotation curve, we radially interpolate the parameters ($v_{\text{sys}}$, PA, $\phi$) to generate the model for the 2D velocity field, which radially traces the rotation curve of the host galaxy stellar component. 

\begin{figure*}
    \hfill\includegraphics[width=\textwidth]{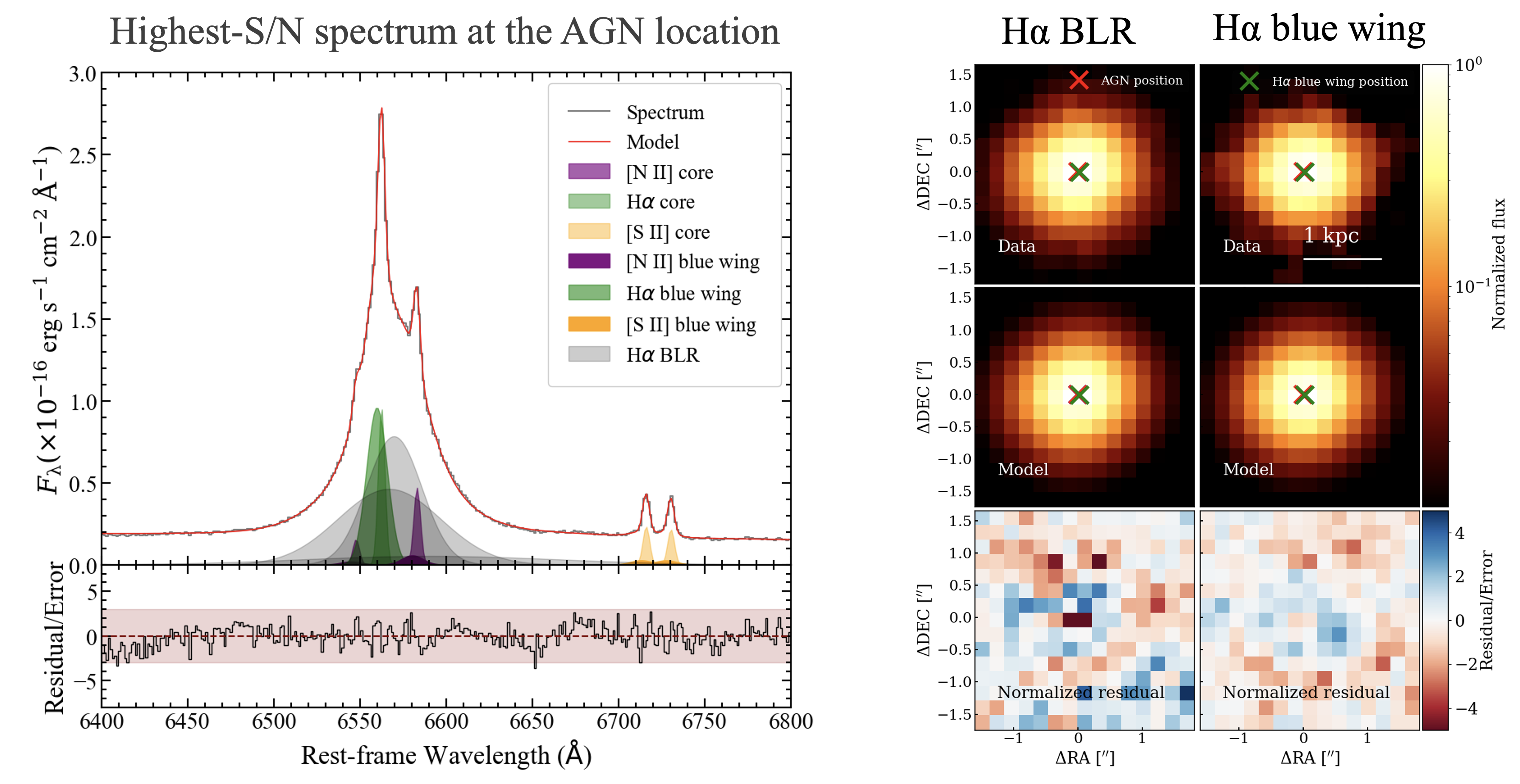}\hspace*{\fill}
    \caption{Spectro-astrometric analysis of the H$\alpha$ emission within the central $3\arcsec \times 3\arcsec$ diameter region. The upper portion of the left panel shows the $3\arcsec \times 3\arcsec$ spectrum (black continuous line) together with the multi-component fit of the emission line spectrum (red continuous line) in the H$\alpha$ + [\ion{N}{2}]~$\lambda\lambda 6548, 6583$ window. The Gaussian components denoting H$\alpha$ NLR emission are shown via the green-shaded regions. The purple-shaded Gaussian components represent the [\ion{N}{2}]~$\lambda\lambda 6548, 6583$ emission lines, and the orange-shaded Gaussian components describe the [\ion{S}{2}]~$\lambda\lambda 6716, 6731$ emission line doublets. We adopt light-colored shades for the narrow core and dark shades for the blue wing components. The gray-shaded Gaussian components define three BLR Gaussians $B_{0}$, $B_{1}$ and $B_{2}$. The lower portion of the left panel shows the normalized residual spectrum (black line) which is contained in the $3\sigma$ region (shaded red) across the entire wavelength range, representing the residuals divided by the error spectra. The corresponding 2D light profiles $\Sigma_{\text{2D}}$ of the H$\alpha$ broad and wing component are shown in the right panels. From top to bottom, the panels describe the  measured $\Sigma_{\text{2D}}$ profile, the best-fit Moffat model and the residual maps normalized by the uncertainty. The red cross describes the AGN location, whereas the green cross highlights the flux-weighted centroid of $\Sigma_{\text{2D}}$, i.e., the component's location. The corresponding error-normalized residual maps show no systematic sub-structures, implying that both the H$\alpha$ broad and wing components are spatially unresolved.}
    \label{fig:O1}
\end{figure*}

In order to achieve a minimum $S/N \sim 20$ of the stellar emission, we use the Voronoi tessellation technique \citep{Cappellari2003} and co-add spectra within cells. The stellar velocity field shows a smooth rotation curve and is largely unperturbed with a median $\text{PA} = 69 \pm 5\arcdeg$. The rotation curve increases linearly from $4~\text{km s}^{-1}$ near the nucleus to $43~\text{km s}^{-1}$ at $4\arcsec$ ($3.2~\text{kpc}$) where it flattens. However, we note that the velocity near the center should be regarded as a lower limit due to beam smearing. Based on the best fit model, we spatially map the velocity field with respect to the kinematic center of the disk-like rotation, that is HE~0040$-$1105's systemic velocity $v_{\text{sys}} = cz = 12583.2 \pm 0.5~\text{km s}^{-1}$, in Figure~\ref{fig:O1}. HE~0040$-$1105's median stellar velocity dispersion is $\sigma_{*} = 128 \pm 2~\text{km s}^{-1}$.

\subsection{The nuclear ionized gas outflow O1}\label{subsec:ionized_gas_outflow}

An important feature that has already been identified in \citet{Singha2021a} is the warm ionized gas outflow driven by the AGN with a size of $741 \pm 3~\text{mas}$. Although this outflow was identified from the H$\beta$ + [\ion{O}{3}]~$\lambda\lambda 4959, 5007$ + \ion{Fe}{2}~$\lambda\lambda 4923, 5018$ emission line complex, it has a corresponding component in the H$\alpha$ emission that has not yet been analyzed.

We first constrain this component's location with a similar approach as described in \citet{Singha2021a} before we estimate the outflow integrated energetics and discuss its impact on the host galaxy.

\subsubsection{Identifying the outflowing H$\alpha$ component}\label{subsubsec:identifying_Ha_outflow}

We constrain the outflow properties by modelling its emission line spectrum with a superposition of multiple Gaussian components. For the AGN continuum, we use a linear model, which provides a sufficient description over the relatively narrow wavelength range analyzed. The broad component of the H$\alpha$ emission line is considerably more complex than that of H$\beta$, so we require three components $B_{0}$, $B_{1}$, $B_{2}$ to describe the broad emission line profile.

The narrow H$\alpha$ + [\ion{N}{2}]~$\lambda\lambda 6548, 6583$ + [\ion{S}{2}]~$\lambda\lambda 6716, 6731$ emission lines are robustly described by a linear superposition of kinematically-coupled Gaussian components. We keep the line flux ratios for [\ion{N}{2}]~$\lambda 6548$/[\ion{N}{2}]~$\lambda 6583$ tied to their theoretical prediction of $1/3$ \citep{Storey2000}. To account for the outflow that manifests as a blue wing, we describe each of the lines with two Gaussian components -- a first narrow `core' component that describes the ionized gas locally to the galaxy and a second broader `wing' component that represents the ionized gas outflow. These constraints allow us to achieve a robust fit for the emission lines.

We fit the spectrum extracted from the central brightest pixel using a non-linear Levenberg-Marquardt algorithm and estimate the uncertainties by performing the fit for 100 mock spectra generated from the corresponding error spectrum.

To estimate if multiple components are required, we use the Akaike information criterion \citep[AIC;][]{Akaike1974} where we penalize the additional component if the value of $\text{AIC}_{\text{s}} = 2 \ln{(L_{\text{s}})} + 2 k_{\text{fs}}$ is greater than 40. Here, $L_{\text{s}}$ represents the likelihood of the model and $k_{\text{fs}}$ indicates the number of free parameters used by the model. Furthermore, we only adopt the additional outflow component if its $S/N$ is greater than 5.

For the best-fit model, we find that the outflow component is relatively broad. Its width as quantified by the velocity range that contains $80\%$ of the line flux is $W_{80,~\text{H}\alpha,~\text{wing}} > 400~\text{km s}^{-1}$, indicating that the outflow is driven by the AGN \citep{Harrison2014,McElroy2015}.

\subsubsection{Size of the ionized gas outflow}\label{subsubsec:outflow_size}

To spatially trace the 2D light distribution for the different kinematic components present in the complex H$\alpha$ + [\ion{N}{2}] + [\ion{S}{2}] complex, we use a similar spectroastrometric approach as described in \citet{Singha2021a}. We fix the kinematic parameters of the emission lines to those retrieved from the central brightest spectrum while only varying the line flux across the FoV. A significant difference from \citet{Singha2021a}, however, is that for the multiple broad components, we only keep the single-line flux of $B_{1}$ as a free parameter while keeping the line flux ratios $B_{0}/B_{1}$ and $B_{2}/B_{1}$ fixed to what we derived from the central spectrum. Fixing the line ratios is a sensible approach since the emission from the BLR is spatially unresolved.

After fitting each spaxel in the inner $3\arcsec \times 3\arcsec$ with the compound model, we construct the 2D surface brightness profile $\Sigma_{\text{2D}}$ for each of the individual emission lines. Since the high $S/N$ broad emission is spatially unresolved, $\Sigma_{\text{2D}}$ corresponds to the empirical point spread function (PSF). We then fit the empirical PSF with a 2D Moffat model, as well as $\Sigma_{\text{2D}}$ of the H$\alpha$ blue wing ($\Sigma_{\text{2D, H}\alpha, ~\text{wing}}$). Following the procedures described in \citet{Singha2021a}, we find that the Moffat model well describes the $\Sigma_{\text{2D, H}\alpha,~\text{wing}}$ profile, suggesting that the blue wing in H$\alpha$ is spatially unresolved by MUSE. The empirical PSF extracted at H$\alpha$ differs ($\text{FWHM} = 656~\text{mas}$) from that at H$\beta$ ($\text{FWHM} = 741~\text{mas}$) because the line shape of the PSF varies with wavelength \citep{Cypriano2010,Eriksen2018}.

\begin{figure}
    \hfill\includegraphics[width=\columnwidth]{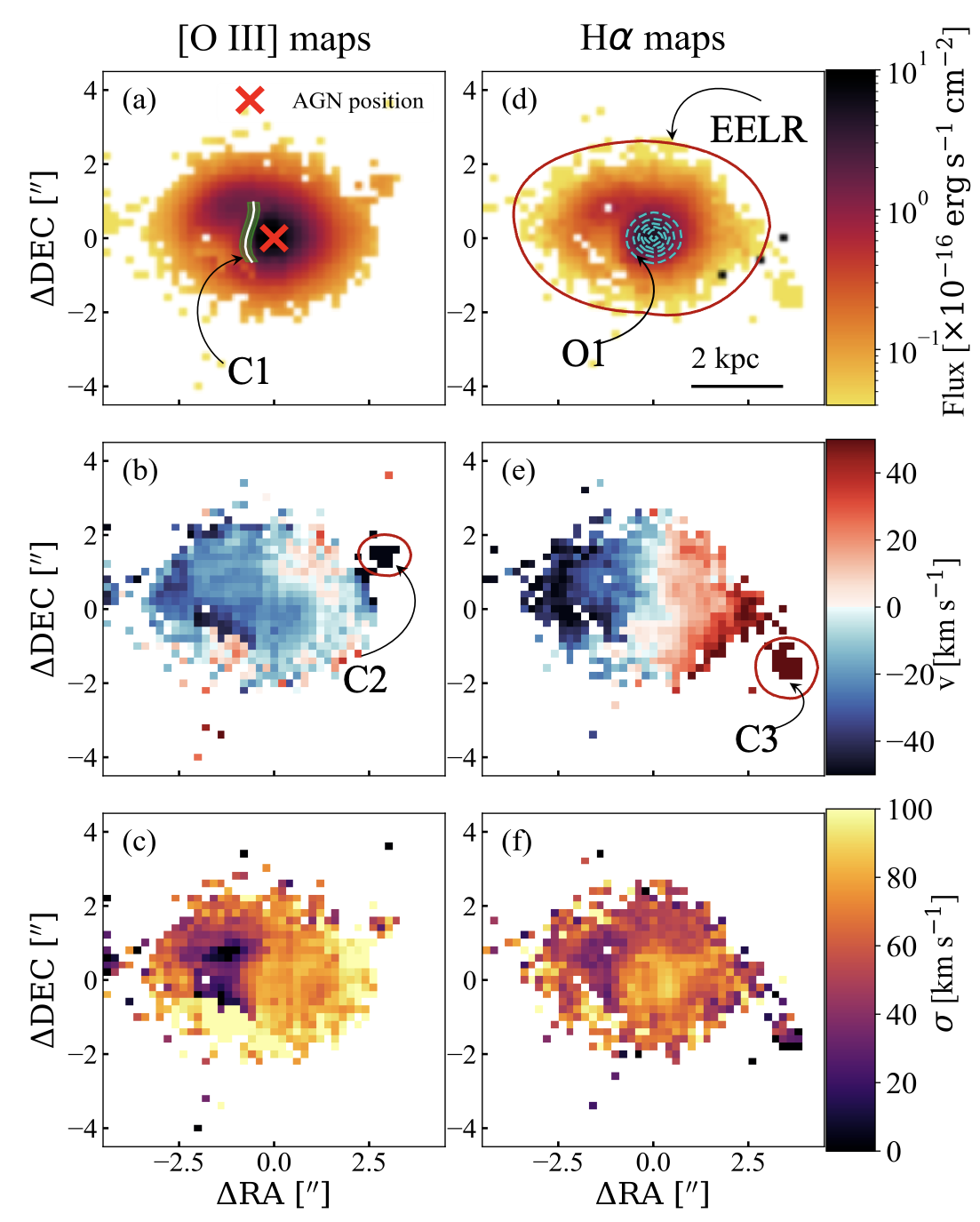}\hspace*{\fill}
    \caption{Mapping HE~0040$-$1105's ionized gas flux and kinematics across the host galaxy. The left panels show the maps extracted from the single-component [\ion{O}{3}] model, and the right panels are those for H$\alpha$. From top to bottom the panels show the emission line flux, rest-frame velocity $v$, and velocity dispersion $\sigma$, respectively. The EELR that is local to the galaxy (see Section~\ref{subsection:energetics_comparison}) is highlighted by the red contour in panel (d). The contours of ionized gas outflow in the center (see Section~\ref{subsec:ionized_gas_outflow}) are shown as dashed cyan lines. In addition, we highlight the contours of the kinematic features analyzed in Section~\ref{subsec:resolved_host_galaxy_emission} where the red line in panel (b) describes the morphology of the receding shell C1, and the red line in panel (e) indicates the H$\alpha$ emitting receding region C3. Both [\ion{O}{3}] emission line nebulae have similar morphologies. However, while the H$\alpha$ velocity field has a clear rotational pattern and a nucleated peak of the velocity dispersion, the [\ion{O}{3}] shows a more chaotic motion with a velocity dispersion in C1 that is lower than average.}
    \label{fig:gas_maps}
\end{figure}

We estimate the statistical uncertainty of the astrometrically measured quantities using a Monte Carlo approach, similar to what is described in \citet{Singha2021a}. We adopt the standard deviation of the distribution as the statistical uncertainty of the centroid location $\sigma_{\text{stat}}$. Furthermore, we estimate the systematic uncertainty of our spectro-astrometric measurements $\sigma_{\text{sys, cor}}$ as described in \citet{Singha2021a}. In this way, we account for systematic effects on our 3D spectroscopic data set, including detector noise and the geometric distortion of the CCD. The total uncertainty adopted for the following analysis is $\sigma_{\text{cor}} = (\sigma_{\text{stat}}^2 + \sigma_{\text{sys}}^2)^{1/2}$.

The offsets between the flux-weighted centroids of the [\ion{O}{3}] and H$\alpha$ blue wings and the AGN nucleus in the kinematically perturbed region O1 (see Figure~\ref{fig:gas_maps}) are \mbox{$d_{\text{AGN}}^{\text{[O~III], O1}} = 92 \pm 8~\text{pc}$} and \mbox{$d_{\text{AGN}}^{\text{H}\alpha \text{, O1}} = 29 \pm 7~\text{pc}$}. Constraining the size of this region is limited by the finite width of the MUSE PSF. Since their $\Sigma_{\text{2D}}$ profiles are not resolved by MUSE, we estimate $\text{FWHM}^{\text{maj}}_{\text{MUSE}}$, where we adopt the width of the elliptical PSF along the major axis as $\text{FWHM}^{\text{maj}}_{\text{MUSE}} = 741 \pm 3~\text{mas}$ (for H$\beta$) and $656 \pm 3~\text{mas}$ \citep[for H$\alpha$;][]{Singha2021a}. If we assume that both [\ion{O}{3}] and H$\alpha$ blue wings originate from the same intrinsic outflow O1, it is sensible we adopt the larger size of the PSF widths, that is the one measured at H$\beta$, to comprehend the maximum possible size of the region perturbed by the outflow. In coordinates of the galaxy system, the maximum projected radial distance of the outflow from the nucleus amounts to $308~\text{pc}$.

We note that the derived offsets and sizes are prone to projection effects. The deprojected offsets could only be calculated if the inclination angle of the AGN nucleus $i$ of HE~0040$-$1105 with respect to our line-of-sight is known. Since the HE~0040$-$1105 outflow inclination is unconstrained, we can use the statistical estimate of the mean unobscured AGN inclination described in \citet{Singha2021a}, $i_{\text{mean}} = 40 \pm 2 \arcdeg$, which results in a maximum intrinsic outflow radius, $d_{\text{max, intr}} = 308~\text{pc}/\text{sin}(i) = 480~\text{pc}$.

For the following analysis, unless stated differently, we adopt the projected distances which involve fewer assumptions on the geometry.

\subsubsection{Outflow energetics}\label{subsubsec:outflow_energetics}

\begin{deluxetable*}{cCCCCCC}
    \tablecaption{Derived parameters for the energetics of HE~0040$-$1105's ionized gas outflow O1. We compute the outflowing mass rate $\dot{M}_{\text{out}}$, the momentum injection rate $\dot{P}_{\text{out}}$ and the kinetic energy injection rate $\dot{E}_{\text{out}}$ for a cone geometry (left columns) and a shell-like geometry (right columns). The rows contain the values for the different parameterizations of the velocity. While the geometry only affects the $0.2~\text{dex}$, the choice of the velocity can change the momentum and energy loading by more than one order of magnitude.\label{tab:energetics}}
    \tablehead{& \multicolumn{3}{c}{Cone geometry} & \multicolumn{3}{c}{Shell geometry} \\ & \colhead{$\log(\dot{M}_{\text{out}}/\text{M}_{\odot}~\text{yr}^{-1})$} & \colhead{$\log(\dot{P}_{\text{out}}/\text{dyne})$} & \colhead{$\log(\dot{E}_{\text{out}}/\text{erg s}^{-1})$} & $\log(\dot{M}_{\text{out}}/\text{M}_{\odot}~\text{yr}^{-1})$ & $\log(\dot{P}_{\text{out}}/\text{dyne})$ & $\log(\dot{E}_{\text{out}}/\text{erg s}^{-1})$}
    \startdata
    $v_{\text{broad}}$ & 0.66 & 33.72 & 40.68 & 0.86 & 33.92 & 40.88 \\
    $W_{80}/1.3$ & 0.92 & 34.24 & 41.46 & 1.12 & 34.44 & 41.66 \\
    $v_{\text{max}}$ & 1.23 & 34.86 & 42.39 & 1.43 & 35.06 & 42.59 \\
    \enddata
    \tablecomments{We have not provided the uncertainties with these measurements as they are smaller than the scatter derived using} different assumptions. The assumed velocity values are $v_{\text{broad}} = -183~\text{km s}^{-1}$, $W_{80}/1.3 = 332~\text{km s}^{-1}$ and $v_{\text{max}} = 680~\text{km s}^{-1}$.
\end{deluxetable*}

We estimate the outflow energetics from the integrated flux from the ionized gas outflow O1. As a first step, we determine the outflowing ionized gas mass assuming `Case B' recombination with the Balmer decrement $\text{H}\alpha / \text{H} \beta = 2.86$ and an electron temperature of $T_{\text{e}} \approx 10^{4}~\text{K}$. Following \citet{Harrison2014}, the ionized gas mass can be estimated as
\begin{equation}
    \frac{M_{\text{ion}}}{2.82 \times 10^{9}~\text{M}_{\odot}} = \left( \frac{L_{\text{H}\beta,~\text{O1}}}{10^{43}~\text{erg~s}^{-1}} \right) \left( \frac{n_{\text{e, O1}}}{100~\text{cm}^{-3}} \right)^{-1},
\end{equation}
where $L_{\text{H}\beta,~\text{O1}}$ is the extinction-corrected luminosity in H$\beta$ and $n_{\text{e, O1}}$ the electron density for O1. We infer a line-of-sight attenuation from a Milky Way-like attenuation curve, following \citet{Cardelli1989}. We estimate $A_{\text{V,~O1}} = 2.71 \pm 0.68~\text{mag}$, indicating significant dust extinction in the system O1. The extinction uncorrected H$\beta$ luminosity in O1 is $\log{(L_{\text{H}\beta,~\text{O1, obs}}/\text{erg s}^{-1})} = 40.13 \pm 0.01$. We correct $L_{\text{H}\beta,~\text{O1, obs}}$ with the optical extinction and estimate the extinction corrected H$\beta$ luminosity as
\begin{equation}
    L_{\text{H}\beta,~\text{O1, int}} = L_{\text{H}\beta,~\text{O1, obs}} \times 10^{A_{\text{V, O1}}/2.5}.
\end{equation}

$\log{(L_{\text{H}\beta,~\text{O1, int}}/\text{erg~s}^{-1}})$ is estimated to be $41.21 \pm 0.27$, which yields $M_{\text{out}} = (3.96 \pm 0.07) \times 10^{6}~\text{M}_{\odot}$.

The derived energetics values depend on the assumed geometry, which is unresolved in our IFU observations. To estimate the outflow energetics we therefore adopt geometries discussed in previous studies. A bi-conical outflow geometry has been used in several studies \citep[e.g.,][]{canodiaz2012,cresci2015,Fiore2017} where the cones are homogeneously filled with gas. In this case, the mass outflow rate can be estimated from the outflow velocity $v_{\text{out}}$ as
\begin{equation}
    \frac{\dot{M}_{\text{ion, cone}}}{\text{M}_{\odot}~\text{yr}^{-1}} = 3 \left( \frac{v_{\text{out}}}{100~\text{km s}^{-1}} \right) \left( \frac{M_{\text{ion}}}{10^{7}~\text{M}_{\odot}} \right) \left( \frac{R_{\text{out}}}{\text{kpc}} \right)^{-1},
\end{equation}
where $R_{\text{out}}$ indicates the distance from the nucleus for which we adopt $d_{\text{max, intr}} = 480~\text{pc}$, the deprojected maximum radius of the ionized gas outflow (see Section~\ref{subsubsec:outflow_size}). We further estimate the momentum injection rate as $\dot{P}_{\text{ion}} = \dot{M}_{\text{ion}} v_{\text{out}}$ and the kinetic energy injection rate as $\dot{E}_{\text{ion}} = \dot{M}_{\text{ion}}v^{2}/2$.

While the choice of $v_{\text{out}}$ significantly affects $\dot{M}_{\text{ion,~cone}}$, past studies have assumed different parameterizations such as the velocity range $W_{80}/1.3$ \citep{Rupke2013TheMergers,Harrison2015,McElroy2015,Husemann2019}, the radial velocity of the blue-wing component $v_{\text{broad}}$, or the maximum velocity of the emission line gas, $v_{\text{max}}$. To account for the systematic differences between the prescriptions, we compute the outflow energetics for each of them.

\begin{figure}
    \includegraphics[width=\columnwidth]{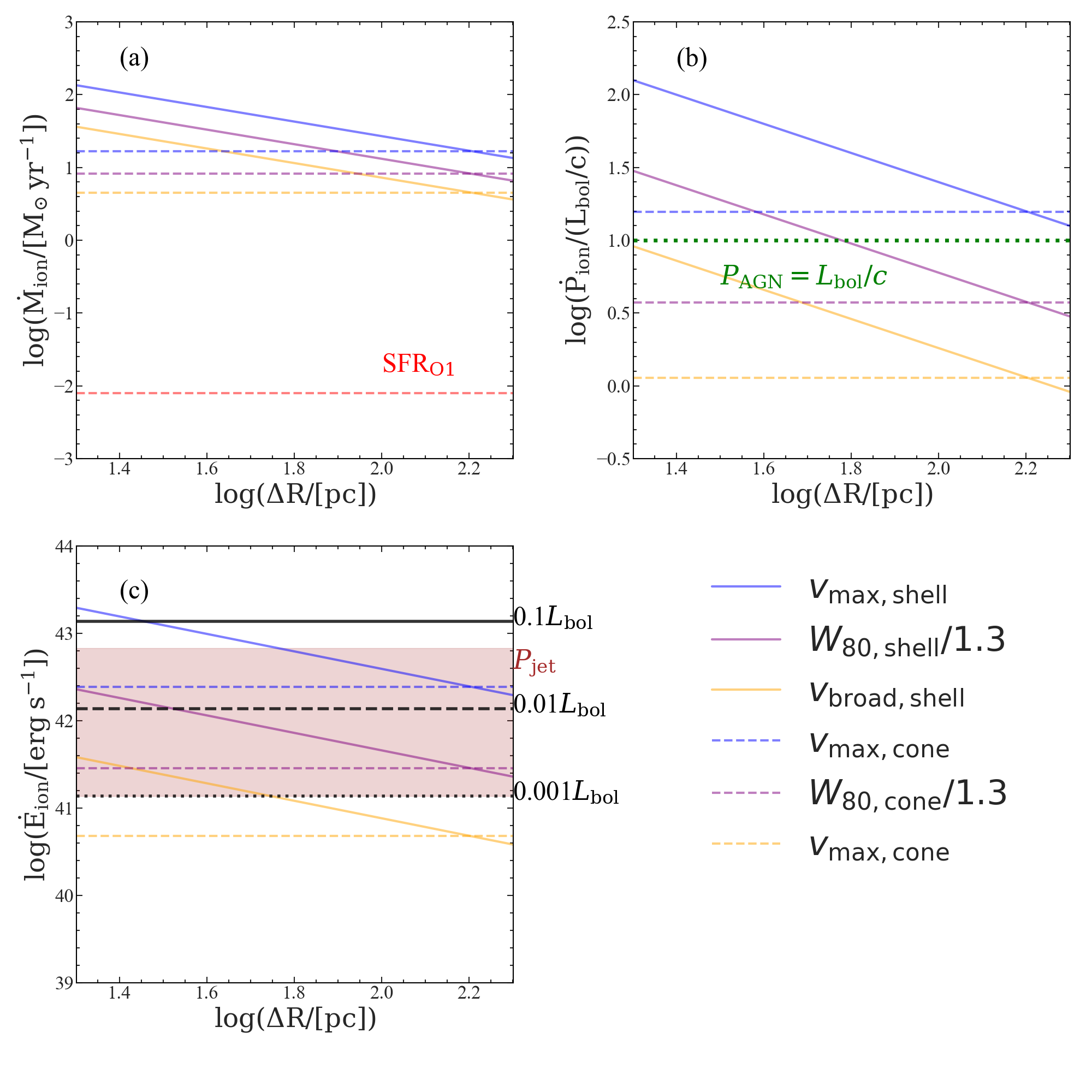}
    \caption{Outflow energetics parameters versus the shell-thickness for different $v_{\text{out}}$ and different outflow geometries. (a) Mass outflow rates, (b) momentum injection rates, and (c) kinetic energy injection rates. The blue, purple and yellow lines correspond to $v_{\text{out}} = v_{\text{max}}$, $W_{80}/1.3$, and $v_{\text{broad}}$ respectively. We use solid lines to showcase the parameters derived for the shell geometry and dashed lines for the cone geometry. The red dashed line in (a) corresponds to the H$\alpha$ derived nuclear SFR. The green dotted line in (b) shows the momentum injection by the AGN radiation field, $P_{\text{AGN}} = L_{\text{bol}}/c$. The black lines in (c) show the $0.1\%$ (dotted line), $1\%$ (dashed line), and $10\%$ (solid line) energy released from the AGN, respectively. The energy-driven mechanism is feasible. The brown shaded region denotes the possible values of the jet-power $P_{\text{jet}}$.}
    \label{fig:energetics}
\end{figure}

To test the impact of the assumed outflow geometry on the derived quantities, we now change to a shell-like geometric model. In this case, the mass outflow rate can be estimated following \citet{Husemann2019}:
\begin{equation}
    \frac{\dot{M}_{\text{out,~shell}}}{\text{M}_{\odot}~\text{yr}^{-1}} = \left( \frac{v_{\text{out}}}{100~\text{km s}^{-1}} \right) \left( \frac{M_{\text{out}}}{10^{6}~\text{M}_{\odot}} \right) \left( \frac{100~\text{pc}}{\Delta R} \right),
\end{equation}
where $\Delta R$ is the thickness of the shell. \citet{Husemann2019} estimated $\Delta R = 20-500~\text{pc}$ for the shell thickness of an ionized gas outflow of $1~\text{kpc}$ extent. For HE~0040$-$1105, we have shown in \citet{Singha2021a} that the outflow is located at less than $< 500~\text{pc}$ from the nucleus. Therefore, we show the injection rates of mass, momentum and kinetic energy for $\Delta R = 20-250~\text{pc}$ (reducing the upper limit on $\Delta R$ by a factor of two) in Figure~\ref{fig:energetics}, and use $\Delta R = 100~\text{pc}$ in Table~\ref{tab:energetics}.

The estimated $\dot{M}_{\text{ion}}$ values for both geometries and for all velocities are at least an order of magnitude higher than those predicted from the AGN-wind scaling relations for ionized gas outflows by \citet{Fiore2017}. The lower limit of $\dot{M}_{\text{ion}}$ is close to the $\log{(L_{\text{bol}})} - \dot{M}_{\text{ion}}$ relation for the luminous quasars as shown in \citet{Singha2021}. The upper limit of $\dot{M}_{\text{ion}}$ in HE~0040$-$1105 is about an order of magnitude higher than that expected from the scaling relation. However, its value is consistent with the $\dot{M}_{\text{ion}}$ of LERGs at similar $L_{\text{bol}}$, which are systems in which the radio source increases the mass loading \citep{Singha2021}.

\subsubsection{Star formation rate in O1}\label{subsubsec:SFR_outflow}

Star formation-related processes can inject momentum and energy into the ambient medium, which may be able to explain the observed ionized gas outflow signatures on several hundred-pc scales. We now aim to estimate the integrated star formation rate (SFR) over the central $1\arcsec$ aperture, to estimate whether nuclear star formation can drive the ionized gas outflow. Using the BPT diagnostic we conclude that HE~0040$-$1105's warm gas component is predominantly ionized by the AGN \citep[][]{Irina2021}. From the H$\alpha$ luminosity, we can therefore only estimate an upper limit for star formation. We first correct the H$\alpha$ luminosity for dust extinction in the galaxy following \citet{Winkel:2022a} which yields an extinction-corrected luminosity of $\log{(L_{\text{int, H}\alpha}/\text{erg~s}^{-1}}) = 39.18 \pm 0.17$. Then, we use the calibration of \citet{Calzetti2007} to convert it into a star formation rate:
\begin{equation}
    \frac{\text{SFR}}{\text{M}_{\odot}~\text{yr}^{-1}} = 5.3 \times 10^{-42} \left( \frac{L_{\text{int, H}\alpha}}{\text{erg s}^{-1}} \right),
\end{equation}
where we find an upper limit of SFR of only $8 \times 10^{-3}~\text{M}_{\odot}~\text{yr}^{-1}$ within the central $< 1\arcsec$.

\subsection{Spatially resolved ionized gas}\label{subsec:resolved_host_galaxy_emission}

\subsubsection{Single component emission line modelling}\label{subsubsec:ionized_gas_vel_single_component}

In order to analyze the resolved emission of ionized gases, we first subtract the unresolved emission from the original data cube. The unresolved emission consists of the BLR emission together with the spatially unresolved blue-wing components that originate from the outflowing system O1. After subtracting point-like emission, the 3D cubes span $4750 - 5090~\text{\AA}$ and $6400 - 6800~\text{\AA}$ and exclusively contain the unresolved host galaxy emission, including the contribution from the EELR.

We employ a simple single-component model to fit the emission line shape throughout the MUSE FoV to characterize the emission of HE~0040$-$1105's ionized gas component. Here, we independently model the emission line complexes H$\beta$ + [\ion{O}{3}] and H$\alpha$ + [\ion{N}{2}] + [\ion{S}{2}] using a linear superposition of kinematically-coupled Gaussian components. Similarly to Section~\ref{subsubsec:identifying_Ha_outflow}, we fix the line ratios among the line doublets [\ion{O}{3}]~$\lambda 4959$/[\ion{O}{3}]~$\lambda 5007$ and [\ion{N}{2}]~$\lambda 6548$/[\ion{N}{2}]~$\lambda 6583$ to their theoretical prediction of $1/3$.

We notice that the single-component model does not properly describe the emission line shapes in some regions of HE~0040$-$1105's ionized gas nebula. In regions where the line shape differs, we first constrain the kinematics of this structure by disentangling the low $S/N$ components from the integrated spectra of multiple $3 \times 3$ spaxel apertures. After the kinematics are constrained, all spaxels are then independently fitted with the two-component model, and the AIC criterion described in Section~\ref{subsubsec:identifying_Ha_outflow} determines whether it provides a better description of the emission line shape than the single-component model. We find that the single-component model has a contribution across the entire FoV such that C2 corresponds to the local EELR of HE~0040$-$1105's host galaxy.

From both the spatial distribution of the ionized gas flux and the kinematic components present in their emission lines, we identify three additional features (C1, C2, and C3; see Figure~\ref{fig:gas_maps}) that are not associated with the local host galaxy EELR. Their kinematic properties, locations, and sizes are listed in Table~\ref{tab:gas_cloud_properties}. In the following, we will characterize them, along with the EELR, in more detail.

\begin{deluxetable*}{ccCCCCccc}
    \tablecaption{Kinematic features identified in HE~0040$-$1105's ionized gas nebula and their properties. From left to right the columns contain the feature name, the emission line in which it is detected, the kinematic parameters velocity $v$, dispersion $\sigma$, line width $W_{80}$, and peak velocity $v_{\text{max}}$, along with a distance to the central AGN $d_{\text{AGN}}$, whether the feature is resolved or not, and the projected spatial size.\label{tab:gas_cloud_properties}}
    \tablehead{\colhead{Name} & \colhead{Emission lines} & \colhead{$v$} &\colhead{$\sigma$} &\colhead{$W_{80}$} & \colhead{$v_{\text{max}}$} & \colhead{$d_{\text{AGN}}$} 
    & \colhead{resolved} & \colhead{size\tablenotemark} \\[-0.20cm] & & \colhead{[$\text{km s}^{-1}$]} & \colhead{[$\text{km s}^{-1}$]} & \colhead{[$\text{km s}^{-1}$]} & \colhead{[$\text{km s}^{-1}$]} & \colhead{[$\text{kpc}$]} & & \colhead{[$\text{kpc}$]}}
    \startdata
    \multirow{2}{*}{O1} & H$\beta$ + [\ion{O}{3}] & -183 $\pm$ 4 & 249 $\pm$ 3 & 432 $\pm$ 6 & 680 $\pm$ 7 & $0.092 \pm 0.008$ & \multirow{2}{*}{yes} & \multirow{2}{*}{$< 0.307$} \\
    & H$\alpha$ + [\ion{N}{2}] + [\ion{S}{2}] & -135 $\pm$ 10 & 283 $\pm$ 12 & 566$\pm$ 21 & 700 $\pm$ 26 & $0.028 \pm 0.008$ & \\ \hline
    \multirow{2}{*}{C1} & H$\beta$ + [\ion{O}{3}] & +40 $\pm$ 2 & 132 $\pm$ 2 & 293 $\pm$ 7 & 304 $\pm$ 4 & \multirow{2}{*}{$0.515$} & \multirow{2}{*}{yes} & \multirow{2}{*}{$< 1.5 \times 0.26$} \\
    & H$\alpha$ + [\ion{N}{2}] + [\ion{S}{2}] & + 69 $\pm$ 31 & 116 $\pm$ 15 & 278 $\pm$ 26 & 301 $\pm$ 42 & \\ \hline
    \multirow{2}{*}{EELR} & H$\beta$ + [\ion{O}{3}] & -90 $-$ +60 & 0 $-$ 120 & 0$-$310 & 40 $-$ 320 & \multirow{2}{*}{$-$} & \multirow{2}{*}{yes} & \multirow{2}{*}{$< 2.8 \times 1.8$} \\
    & H$\alpha$ + [\ion{N}{2}] + [\ion{S}{2}] & -60 $-$ +80 & 0 $-$ 120 & 0 $-$ 310 & 10 $-$ 310 & \\ \hline
    \multirow{2}{*}{C2a} & H$\beta$ + [\ion{O}{3}] & -224 $\pm$ 4 & 61 $\pm$ 3 & 156 $\pm$ 8 & 346 $\pm$ 7 & \multirow{2}{*}{$2.6$} 
    & \multirow{2}{*}{no} & \multirow{2}{*}{$< 0.307$} \\
    & H$\alpha$ + [\ion{N}{2}] + [\ion{S}{2}] & -215 $\pm$ 5 & 71 $\pm$ 5 & 181 $\pm$ 35 & 356 $\pm$ 11 \\
    \enddata
\end{deluxetable*}

\subsubsection{The EELR local to the host galaxy}\label{subsubsec:EELR}

The EELR is spatially resolved by MUSE and this large ionized gas nebula extends up to a distance $\sim 2.8~\text{kpc}$ from the central nucleus. In Figure~\ref{fig:gas_maps}, we show the 2D flux and kinematic profiles for the ionized gas in [\ion{O}{3}] and H$\alpha$.

On the northeastern (NE) side of the nucleus, the [\ion{O}{3}] emitting gas clouds appear blue-shifted, with a median radial velocity $v_{[\text{O~III], EELR}} = -25 \pm 3~\text{km s}^{-1}$ that remains uniform up to the outer boundary of the nebula (Figure~\ref{fig:gas_maps}b). We notice that the [\ion{O}{3}] velocity dispersion, $\sigma_{\text{[O~III], EELR}}$ decreases $< 20~\text{km s}^{-1}$ where C1 is located (Figure~\ref{fig:gas_maps}c). On the other hand, the radial velocity ($v_{\text{H}\alpha,~\text{EELR}}$) profile of the H$\alpha$ clouds shows a relatively smooth velocity gradient from the eastern side to the western side (Figure~\ref{fig:gas_maps}e). No significant decrease in their dispersion $\sigma_{\text{H}\alpha,~\text{EELR}}$ is observed at the location of C1, where the median is $\sigma_{\text{H}\alpha,~\text{EELR}} = 35 \pm 7~\text{km s}^{-1}$ (Figure~\ref{fig:gas_maps}f). The gas kinematics in the EELR is thus quiescent, in contrast with O1.

\citet{Nelson1996a} observed a tight correlation between $\sigma_{\text{[O~III]}}$ and $\sigma_{*}$, suggesting that the kinematics of the EELR gas in Seyfert galaxies is controlled by the bulge gravitational potential. Studies by \citet{Greene2005,Bian2006} also found that the median ratios between the velocity dispersion of ionized gas and stars ($r_{\text{ion}/*} = 1.00 \pm 0.35$, and $1.20 \pm 0.96$ respectively) confirm the bulge dominated gravitational motion on kpc scales. The median $r_{\text{ion}/*}$ in the EELR, $r_{\text{ion}/*,~\text{EELR}} = 0.58 \pm 0.22$, implying that the EELR gas in HE~0040$-$1105 is sub-virial, and hence cannot escape the bulge potential of the host.

\subsubsection*{EELR \MakeLowercase{kinematic modeling}}\label{subsubsec:stellar_velocity_field_EELR}

Similar to the stellar velocity field, we fit the kinematic modelling of the 2D radial velocity profile H$\alpha$ ionized gas velocity field with a tilted-ring model as described in Section~\ref{subsec:stellar_velocity_field}. We obtain a robust fit where the small residuals scatter around the rest-frame velocity of the host galaxy. Between the concentric rings, we find a median PA of $16 \pm 2\arcdeg$. The $v_{\text{H}\alpha}$ profile of the EELR is well described by a thin rotating disk model, as shown in Figure~\ref{fig:stellar_vel_model}. We find the PA difference between the stars and the ionized gas, $\text{PA}_{\text{star-gas}} = 53 \pm 2\arcdeg$. The quiescent kinematics of the ionized gas in the EELR and the significant $\text{PA}_{\text{star-gas}}$ are indicative of the external origin of gas in HE~0040$-$1105, such as early, ongoing or late stage mergers \citep[e.g.,][]{Barrera-Ballesteros2015,Jin2016,Davis2011}.

\subsubsection{The receding shell C1}\label{subsubsec:C1}

\begin{figure*}
    \hfill\includegraphics[width=\textwidth]{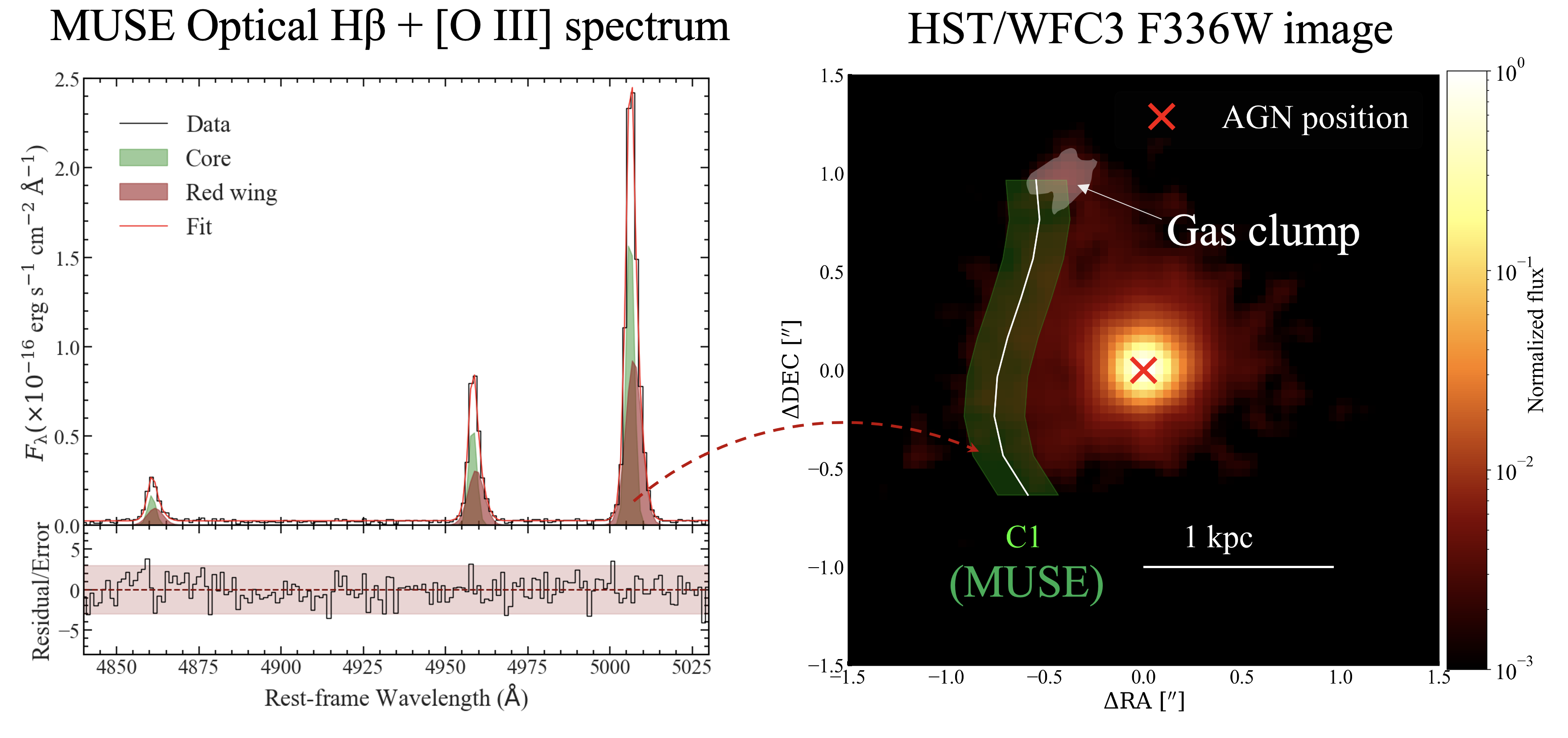}\hspace*{\fill}
    \caption{Multi-wavelength view of the region C1. \textit{Left panel}: MUSE optical spectra extracted from a $3 \times 3$ spaxel region (in gray), along with its multi-Gaussian component fit (in red). The green-shaded Gaussian components represent the narrow core which is a part of the EELR, whereas the dark brown-shaded components trace the red-wing component, corresponding to the region C1. \textit{Right panel}: \textit{HST}/WFC3 near-UV image centered at $3350~\text{\AA}$. The green region describes the area covered by the deconvolved C1, and the white line denotes its centroids. A clumpy structure on the northeastern side of the nucleus is shaded with white to showcase its spatial location and morphology. C1 overlaps with the gas clumps on the north where $S/N > 5$. Such spatial coincidence indicates that C1 is a part of the clumpy region.}
    \label{fig:C1}
\end{figure*}

On the eastern side of the AGN nucleus, a red wing is present in each of the strong emission lines, which is highlighted in Figure~\ref{fig:C1}. From north to south the fitted radial velocities and velocity dispersions are constant across the characteristic with mean values of $v_{\text{red, C1}} = 40 \pm 2~\text{km s}^{-1}$ (H$\beta$, [\ion{O}{3}]) and $\sigma_{\text{red, C1}} = 132 \pm 2~\text{km s}^{-1}$ (H$\alpha$, [\ion{N}{2}], [\ion{S}{2}]) respectively (see Appendix~\ref{appendix:Ha_NII} for the spectral fit). We highlight the spots with maximum luminosities at fixed DEC in Figure~\ref{fig:C1}. The resulting structure appears to be shell-shaped.

\begin{figure*}
    \hfill\includegraphics[width=1.0\textwidth]{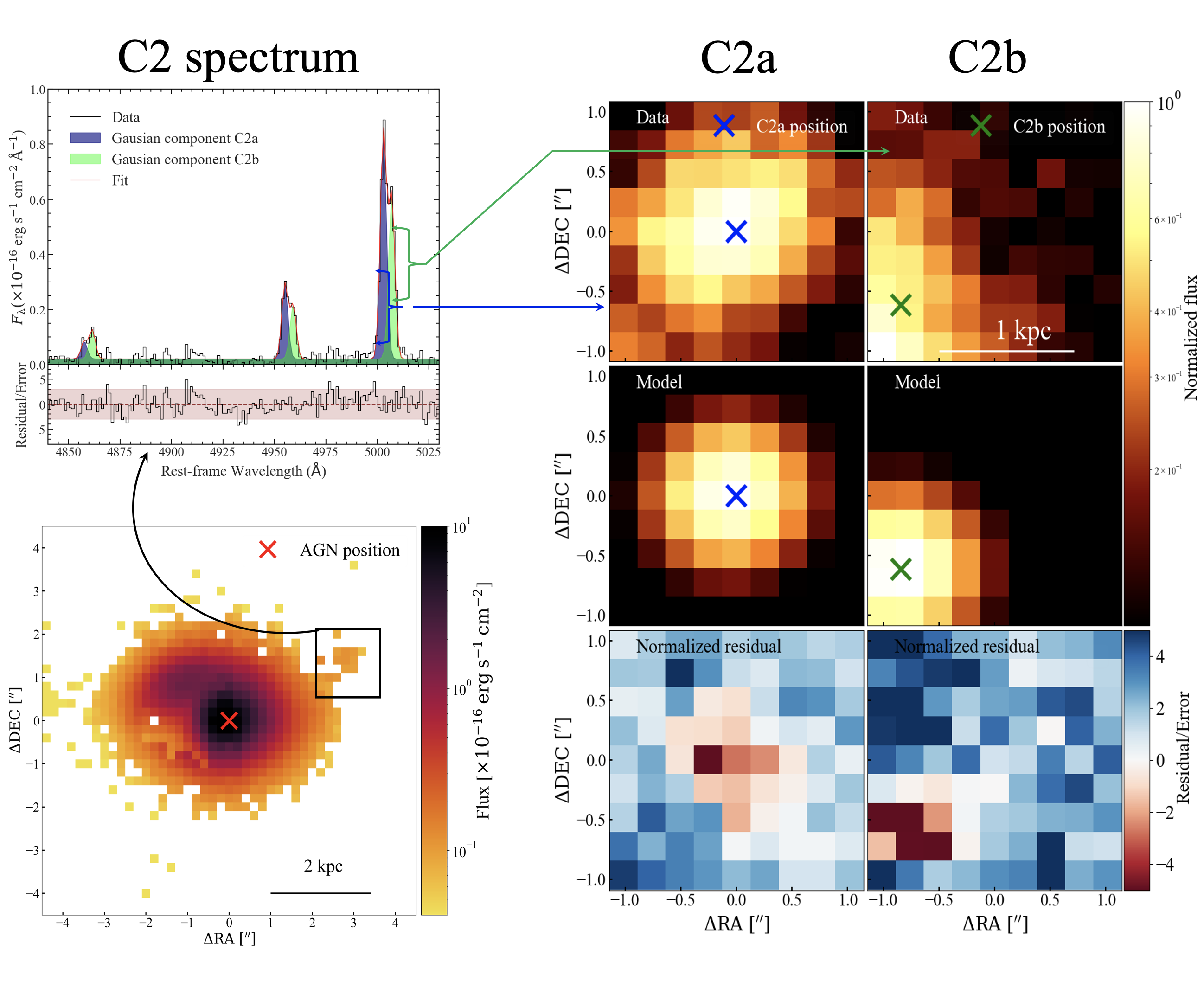}\hspace*{\fill}
    \caption{Spectro-astrometric analysis of the multi-component [\ion{O}{3}] cloud C2. \textit{Lower left panel}: 2D flux profile of [\ion{O}{3}] with C2 being highlighted with a box (upper panel). \textit{Upper left panel}: Optical H$\beta$ + [\ion{O}{3}] spectrum extracted from a $9 \times 9$ spaxel aperture, and its multi-component fit. The gray lines describe the data, and the blue and the green shaded Gaussian components correspond to the regions C2a and C2b respectively (top panel). The residual spectrum is divided by the error spectrum. The brown-shaded region defines the residuals within $3\sigma$ and the red dotted line indicates a region with zero residual (bottom panel). \textit{Right panel}: 2D Moffat modeling results for C2. From top to bottom, the panels show the measured $\Sigma_{\text{2D}}$, the corresponding best-fit 2D Moffat model and the residual maps normalized by the uncertainty. From left to right, the panels correspond to C2a and C2b. The blue and green crosses indicate the flux-weighted centroids of C2a and C2b. The $\Sigma_{\text{2D}}$ profile of C2a is well described by the Moffat PSF, whereas, for C2b, the Moffat model leads to high residuals.}
    \label{fig:knot_maps}
\end{figure*}

\subsubsection*{M\MakeLowercase{orphology of the underlying structure}}

C1 is likely affected by beam smearing, since its emission line kinematics are almost constant across the structure. However, the spatial profile of C1 does not appear to be PSF-like, which could suggest that the observed emission is the result of the PSF convolved with some other underlying emission. Thus, our objective is to obtain spatial information on the possible underlying structure. The detailed procedure of how we extract the underlying structure is described in Appendix~\ref{appendix:deconvolution}. We find that the intrinsic morphology of the gas cloud is similar to a shell, with a distance of  $308~\text{mas}$ ($250~\text{pc}$) from the nucleus.

\subsubsection*{T\MakeLowercase{he clumpy gas in the} UV}

The near ultraviolet image taken with \textit{HST}/WFC3 exhibits structures in HE~0040$-$1105's host galaxy which we detect at the $> 3\sigma$ and $> 5\sigma$ confidence level, respectively. They are located on the outskirts of the PSF, $\sim 800~\text{mas}$ or $660~\text{pc}$ northeast of the AGN location. In Figure~\ref{fig:C1} we highlight the structures together with the deconvolved profile of the shell-like C1 feature, whose locations appear to coincide. Possible explanations include clumpy outflows \citep{Takeuchi2013}, accretion of gas onto the nucleus \citep{Tremblay2016}, and gas clumps due to a merger \citep{Arata2018}. We notice that $\sigma_{\text{red, C1}}$ agrees with $\sigma_{*}$ within uncertainties, implying that these gas clouds could undergo a virial motion influenced by the bulge potential \citep{Nelson1996a} and hence may not be a part of the nuclear outflow.  However, this clumpy gas detected in near-UV is also consistent with HE~0040$-$1105 being a late-stage merger, which is in line with our findings in Section~\ref{subsubsec:EELR}. Therefore the host galaxy of HE~0040$-$1105 is most likely a merger-remnant.

\begin{figure}
    \hspace{-0.2cm}
    \hfill\includegraphics[width=\columnwidth]{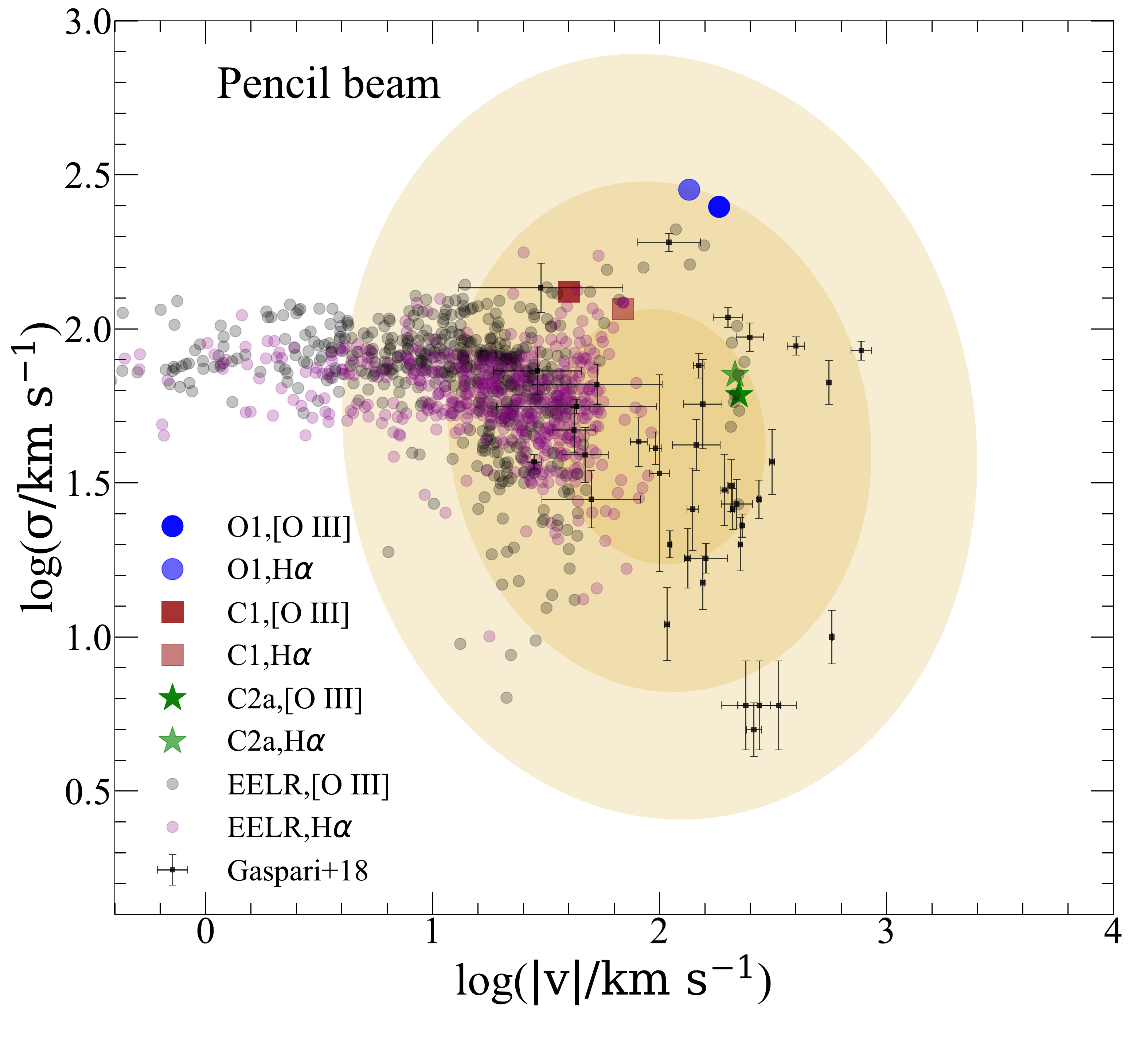}\hspace*{\fill}
    \caption{Kinematical diagnostic (`k-plot') comparing the line broadening versus line shift, with a pencil-beam approach (pixel-by-pixel). The yellow contours show the 1-3$\sigma$ confidence intervals predicted by CCA simulations, while the black points denote a diverse sample of central and isolated galaxies (see \citealt{Gaspari:2018}). We overlay the results for the [\ion{O}{3}] emitting ionized gas in the O1 outflow (blue circles), C1 shell (brown square) and isolated [\ion{O}{3}] cloud C2a (green stars). We adopt similar but light-colored markers to denote the H$\alpha$ emitting gas clouds. The small grey and purple circles denote the ionized [\ion{O}{3}] and H$\alpha$ gas in the EELR. Most gas elements are within the $\sim 2\sigma$ CCA contours suggesting that our main features are likely experiencing turbulent multi-phase condensation, which will soon stimulate a rain back on the SMBH.}
    \label{fig:k-plot}
\end{figure}

\subsubsection{The multi-component cloud C2}\label{subsubsec:C2}

The ionized gas clouds emitting in [\ion{O}{3}] have a radial velocity $v_{\text{[O~III]}} \sim -200~\text{km s}^{-1}$ and dispersion $\sigma_{\text{[O~III]}} \sim 70~\text{km s}^{-1}$ (Figure~\ref{fig:gas_maps}c). To determine the associated gas kinematics, we inspect the emission line spectrum H$\beta$ + [\ion{O}{3}] by co-adding spectra from a $9 \times 9$ spaxel aperture, enabling us to achieve $S/N > 5$ in the emission lines. The bright [\ion{O}{3}] emission line allows one to disentangle two components.  H$\alpha$ + [\ion{N}{2}] + [\ion{S}{2}] emission lines have $S/N < 3$ in each spaxel in C2. In order to infer the ionization condition we co-add spectra from the similar $9 \times 9$ and achieve $S/N>4$, which we fit with a multi-Gaussian model and notice similar double-peaked emission lines (see Appendix~\ref{appendix:Ha_NII}). However, we constrain our analysis to H$\beta$ + [\ion{O}{3}] emission lines due to the availability of higher $S/N$ spectra. The H$\beta$ + [\ion{O}{3}] spectrum contains a double-peaked emission line profile, which we model with a superposition of kinematically-coupled Gaussian components. The AIC criterion suggests that two Gaussian components C2a and C2b for each of the corresponding emission lines are sufficient to describe the shape of the emission line of C2. C2a is blue-shifted by $v_{\text{[O~III], C2a}} \sim -220~\text{km s}^{-1}$, while C2b is red-shifted by $v_{\text{[O~III], C2b}} \sim +30~\text{km s}^{-1}$.

\subsubsection*{O\MakeLowercase{rigin of the} C\MakeLowercase{2 gas clouds}}

We perform a spectro-astrometric analysis within $9 \times 9$ pixels ($1\farcs8 \times 1\farcs8$) to spatially locate C2a and C2b according to \citet{Singha2021a}. The FWHM of the MUSE PSF $>3$ pixels makes any result derived from the analysis within the $3 \times 3$ pixel diameter region unreliable. Choosing a larger aperture produces a much more robust fit. We find that C2a is spatially unresolved by MUSE, whereas C2b appears to be spatially resolved. The projected offset between the flux-weighted centroids of C2b and C2a is $682 \pm 28~\text{pc}$. As seen in Figure~\ref{fig:knot_maps}, the spatial morphology of C2b suggests that it is part of the ionized gas emission from the EELR. The projected offset between the flux-weighted centroid of C2a and the AGN nucleus of HE~0040$-$1105 is $\sim 2.6~\text{kpc}$. Its isolated morphology and sub-virial kinematics (velocity dispersion $< \sigma_{*}$) suggest that the gas in C2a is unlikely to be a part of the nuclear AGN-driven outflow.

\subsubsection{The H$\alpha$-emitting cloud C3}\label{subsubsec:C3}

Another component C3 is located at a large distance from the nucleus ($d \sim 3~\text{kpc}$). As opposed to C2, this feature is only present in H$\alpha$ + [\ion{N}{2}] + [\ion{S}{2}], but not in H$\beta$ + [\ion{O}{3}]. The ionized gas cloud is slightly redshifted with respect to the EELR thin-disk rotation with $\sigma_{\text{H}\alpha}= 40 \pm 2~\text{km s}^{-1}$. However, in Figure~\ref{fig:stellar_vel_model}c, a thin disk model well-predicts the receding line-of-sight velocity on the location of C3, which is further confirmed by the velocity residuals scatter around the rest-frame velocity (Figure~\ref{fig:stellar_vel_model}f). Our results suggest that the thin rotating disk model describes the velocity structure of C3. Since the ionized gas clouds in C3 seem to follow the regular rotation pattern of the EELR, we conclude that C3 is a structure that is embedded in the EELR. The compactness of C3 may therefore be explained by the sensitivity limit of the observation, rather than C3 being an independent gas cloud.

\subsubsection{Chaotic cold accretion and condensation}\label{subsubsec:CCA}

While the motions of the gas are globally affected by the bulge potential in situ, the cloud elements will be also affected by other relevant physical processes. In particular, whenever there are some significant perturbations driven in the gaseous atmospheres, the gas will rapidly condense via turbulent thermal instability \citep[e.g.,][]{Gaspari:2013}. In HE~0040$-$1105, AGN feedback will create fluctuations at small radii. At large scales, the merger environment will further augment the density fluctuations \citep[e.g.,][]{Lau:2017} in the ISM, which will quicken the multi-phase precipitation. To test this, we use the kinematic diagnostic \citep[`k-plot';][]{Gaspari:2018} which confronts the line-of-sight velocity $v$ against the line-of-sight velocity dispersion $\sigma$ (Table~\ref{tab:gas_cloud_properties}), in logarithmic space. This is an insightful diagnostic leveraged in several recent studies \citep[see also][]{Maccagni:2021,North:2021,Temi:2022}.

We show the k-plot in Figure~\ref{fig:k-plot}, superposed to the confidence intervals predicted by high-resolution CCA simulations. Such a diagnostic is very useful here to dissect different physical mechanisms acting on the cold/warm gas. Gas clouds located in the top left quadrant often undergo macro turbulent motions in the galactic/group halo. In contrast, ionized gas with both high velocity and dispersion (top right quadrant) are more likely affected by the outflow kinematics. Cloud elements that reside in the bottom right quadrant are often associated with fast micro-scale inflows falling onto the SMBH. Cloudlets residing in the central $1 - 2 \sigma$ contour regions are likely experiencing the CCA condensation rain, eventually falling back onto the nuclear region.

As shown in Figure~\ref{fig:k-plot}, our HE~0040$-$1105 C1 (red) and C2a (green) points reside within the central $2 \sigma$ CCA (yellow) contour, regardless of the line used. Thus, alongside their moderate velocities, this suggests that C1 and C2a elements are likely prone to multi-phase condensation, and will soon rain back onto the AGN (mainly via the cloud inelastic collisions). The O1 components are instead more elevated, closer to the upper regions, thus suggesting that such gas is still significantly tied to the outflow thrust, yet being capable of moderate condensation. Considering the extended complex of ionized gas, the bulk of the H$\alpha$ (grey) and [\ion{O}{3}] (purple) points still fall within the CCA region, albeit we notice a left straight tail of gas affected by rotation. Overall, on top of pure gravitational dynamics, CCA and related turbulent condensation are likely important physical components acting in HE~0040$-$1105. 

\begin{figure*}
    \centering
    \includegraphics[width=0.8\textwidth]{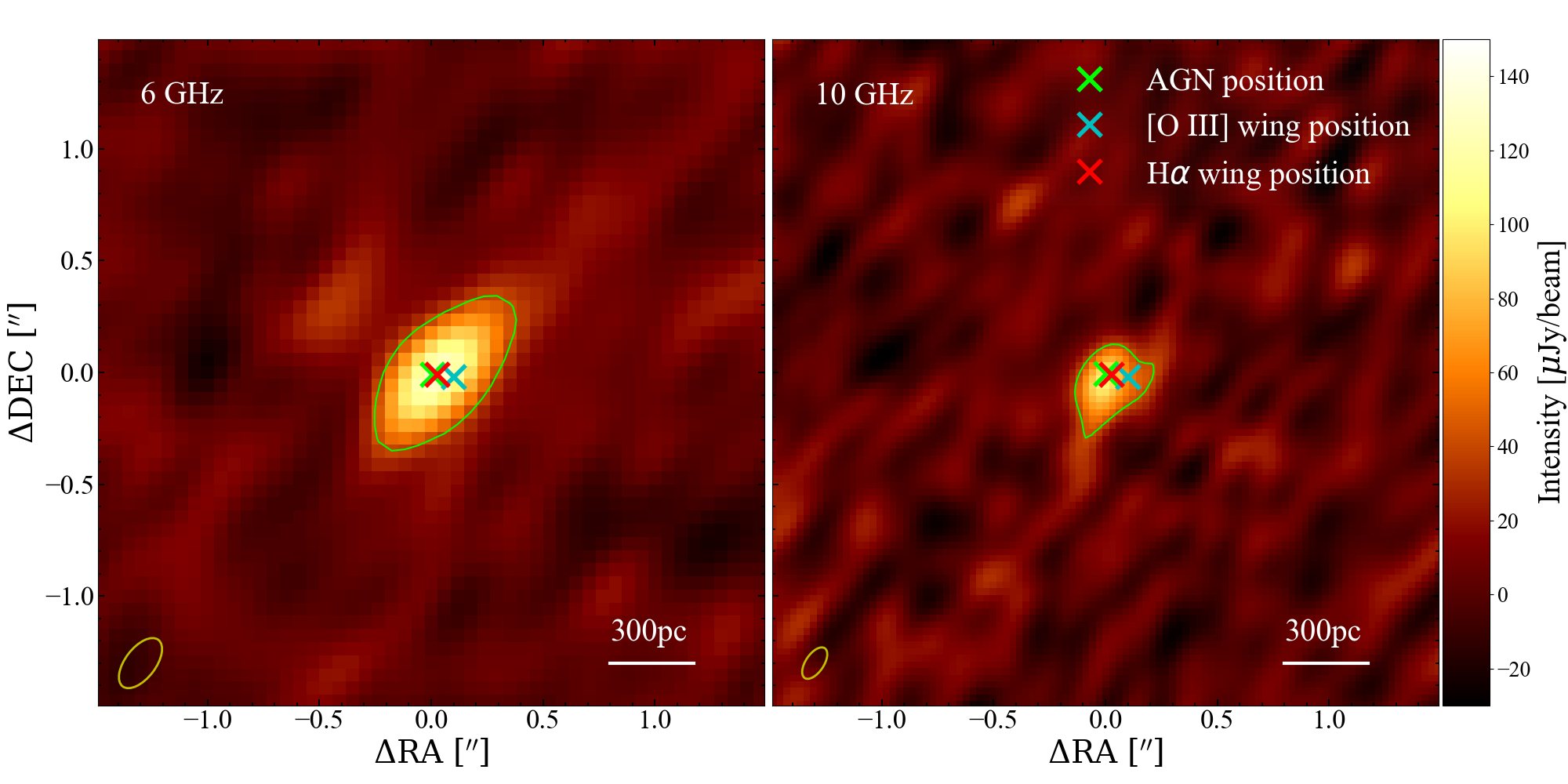}
    \caption{VLA continuum images of HE~0040$-$1105 in the central $3\arcsec$. \textit{Left panel}: The $6~\text{GHz}$ image with the yellow contour in the bottom left indicates the clean beam ($0.53\arcsec \times 0.27\arcsec$ and $\text{PA} = -36.9\arcdeg$). \textit{Right panel}: The $10~\text{GHz}$ image with yellow contour indicating the clean beam ($0.26\arcsec \times 0.17\arcsec$ and $\text{PA} = -22\arcdeg$). Green contours correspond to the $5\sigma$ sensitivity levels at $35~\mu\text{Jy}$ and $40~\mu\text{Jy}$ for the C- and X-band, respectively. The spectro-astrometric positions of the AGN, [\ion{O}{3}] wing, and H$\alpha$ components are highlighted as green, red, and cyan crosses, respectively.}
    \label{fig:MUSE_VLA}
\end{figure*}

\begin{deluxetable*}{cCccCCCCC}
    \tablecaption{Orientation and morphology of the identified radio-emitting regions. From left to right, the columns denote the facility used, the central frequency in the observed frequency band, different regions in the radio emission, spatial morphology denoting if the emission is compact (unresolved) or extended (resolved), the major and minor axes of the de-convolved source from the clean beam, PA of the de-convolved source, the peak brightness, and the integrated flux density. In the case of unresolved radio emission, we tabulate the clean beam size as the upper limits of the underlying radio emission convolved with the clean beam. \textcolor{black}{Archival radio interferometric observations from RACS, and VLASS are listed as well.}
    }\label{table:radio_results}
    \tablehead{\colhead{Facility} & \colhead{Frequency} & \colhead{Region} & \colhead{Morphology} & \colhead{FWHM} & \colhead{PA} & \colhead{Peak brightness} & \colhead{Integrated flux density} \\[-0.20cm]
    & \text{[GHz]} & & & \colhead{[mas]} & \colhead{[$\arcdeg$]} & \colhead{[$\mu$Jy beam$^{-1}$]} & \colhead{[$\mu$Jy]}}
    \startdata
     RACS & 0.88 & - & - & <25000\times25000 & - & 1740\pm340 & 3740\pm690 \\
     EVN & 1.67 & S0 & Unresolved & <20\times10 & - & 83.7 \pm 10.4 & 82 \pm 12 \\
     EVN & 1.67 & S1 & Unresolved & <20\times10 & - & 48.1 \pm 7.8 & 34.0 \pm 8.0 \\
     EVN & 1.67 & S2 & Unresolved & <20\times10 & - & 44.4 \pm 7.6 & 43.0 \pm 9.6 \\
     EVN & 1.67 & S3 & Unresolved & <20\times10 & - & 57.8 \pm 8.4 & 41 \pm12 \\
     VLASS & 3 & - & - & <2500\times2500 & - & 678\pm140 & 1370\pm500 \\
     VLA & 6 & O1 & Resolved & (392 \pm 133) \times (226 \pm 85) & 132 \pm 74 & 125 \pm 10 & 204 \pm 26 \\
     VLA & 10 & O1 & Resolved & (238 \pm 85) \times (163 \pm 94) & 109 \pm 52 & 92 \pm 11 & 171 \pm 28 \\
    \enddata
\end{deluxetable*}

\subsection{Morphology of the radio emission}\label{subsec:radio_emission}

\subsubsection{Sub-kpc scales traced by VLA}\label{subsubsec:VLA}

Our main objective in this section is to compare the O1 outflow to the observed radio emission spatially. We show the radio continuum images of HE~0040$-$1105 at $6$ and $10~\text{GHz}$ obtained using VLA in Figure~\ref{fig:MUSE_VLA}. In order to extract the intensity and morphology of the bright spot in the center, we run the \texttt{imfit} task included in CASA on the VLA C- and X-band radio images. For the measured flux densities, we follow the conservative approach from \citet{Panessa2022} and consider the total uncertainty $\sigma_{S_{\nu}}$ of the peak brightness that involves both the systematic uncertainty from the \texttt{imfit} task, and the statistical uncertainty which equals the sum in quadrature of the off-source rms of the image $\sigma_{\text{off}}$ plus the $5\%$ calibration uncertainty $\epsilon$ times the peak brightness $S_{\nu}$.

We derive the total uncertainty from the quadrature sum of the individual contributions $\sigma_{S_{\nu}} = (\sigma^{2}_{\text{off}} + (\epsilon S_{\nu})^{2})^{1/2}$. The results obtained from our \texttt{imfit} analysis are tabulated in Table~\ref{table:radio_results}. We notice that the major and minor axes of the de-convolved structure in both images (Figure~\ref{fig:MUSE_VLA}) agree within their uncertainties. The similar spatial extensions of the underlying radio emission de-convolved from the clean beam in both $6$ and $10~\text{GHz}$ observations confirm the detection of extended radio emission on $< 500~\text{pc}$ scales. Comparing observations from two different instruments requires a point of reference, for which we chose the AGN location. We assume that the AGN is located at the centroid of the brightest spots in the center of the VLA images or the radio core. We identify the H$\beta$ BLR centroid with the radio core in the VLA image. In Figure~\ref{fig:MUSE_VLA}, we notice that the flux-weighted centroids of [\ion{O}{3}] and H$\alpha$ in O1 overlap with the radio VLA C- ($6~\text{GHz}$) and X- ($10~\text{GHz}$) band continuum emission. Such spatial coincidence suggests that the radio emission detected in VLA is either produced due to the outflows, or the outflows produce the radio emission. However, one more possibility is that the radio structures happen to be co-spatial with the centroids of the outflow without any connection between them.

From our VLA observations, we estimate values of the spectral indices from the peak brightness and the integrated flux density as $\alpha_{\text{peak}} = -0.60 \pm 0.28$ and $\alpha_{\text{int}} = -0.34 \pm 0.41$ respectively. The uncertainties associated with the spectral indices are evaluated as per \citet{Panessa2022}. As the peak brightness is not subjected to the different beam sizes in different frequencies, we use $\alpha_{\text{peak}}$ throughout the paper.

\subsubsection*{P\MakeLowercase{revious radio observations}}

The radio source in HE~0040$-$1105 is undetected in the FIRST survey \citep{Becker1995}, giving an upper limit to the integrated flux density of $1~\text{mJy}$ at $1.4~\text{GHz}$. The source is detected in the VLASS survey \citep{Lacy2020} at $3~\text{GHz}$. The peak brightness and the integrated flux density are $0.678~\text{mJy beam}^{-1}$ and $0.810~\text{mJy}$ respectively. The measured $3-6~\text{GHz}$ spectral index ($\alpha_{\text{3-6 GHz}}$) is steep with $\alpha_{\text{3-6~GHz}} = -2.4$. 
\textcolor{black}{The Rapid ASKAP Continuum Survey (RACS) has also detected the radio source at $880~\text{MHz}$ with a resolution of $25\arcsec \times 25\arcsec$. The peak brightness and flux density values for of this detection are tabulated in Table~\ref{table:radio_results}.}

\subsubsection{Parsec scales resolved by EVN}\label{subsubsec:EVN}

\begin{figure*}
    \centering
    \includegraphics[width=0.8\textwidth]{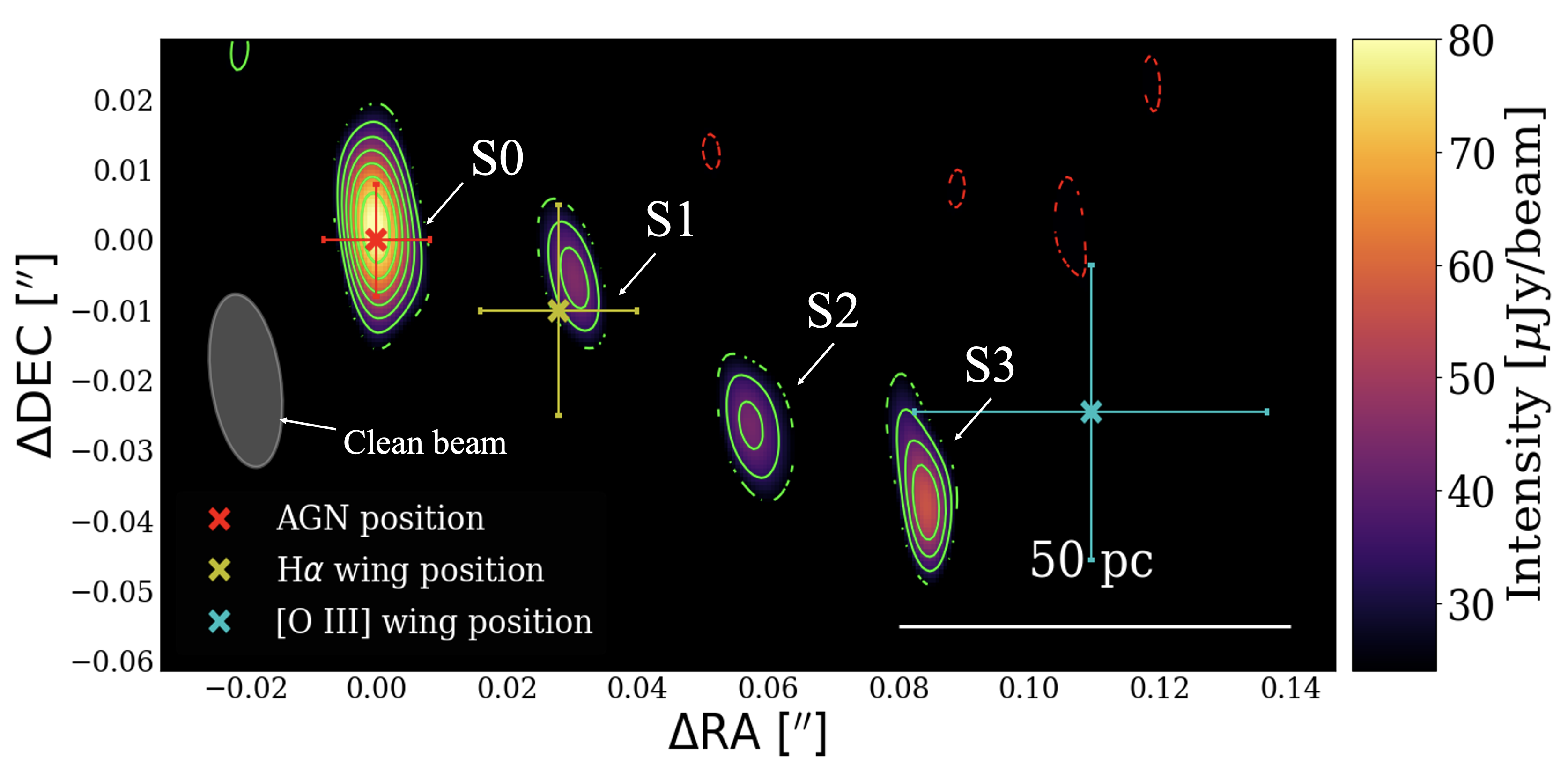}
    \caption{EVN contour image of HE~0040$-$1105 at $18~\text{cm}$. The image shows four unresolved components. We identify the brightest component S0 with the galaxy nucleus and name the remaining components S1, S2, and S3.  The synthesized beam is $24.9 \times 10.7~\text{mas}$ at $8.97\arcdeg$ position angle and is shown on the left edge of the image as a gray-shaded ellipse. The contour levels are -30, 30, 40, 50, 60, 70, 80, and 90 percent of the peak intensity $83.7~\mu\text{Jy beam}^{-1}$. The positive contours are plotted with green lines, whereas the negative contours are indicated with red lines. The crosses describe the flux-weighted centroids of the AGN (red), the H$\alpha$ blue wing (yellow), and the [\ion{O}{3}] blue wing (cyan) emission. The associated error bars indicate the $3 \sigma$ uncertainties of the respective locations determined using spectro-astrometry. Assuming the brightest spot in S0 originates from the AGN, S1 and S3 overlap with the flux-weighted centroids of the blue wings within their uncertainties.}
    \label{fig:EVN}
\end{figure*}

Despite the VLA's high spatial resolution, it cannot resolve the central $200~\text{pc}$ aperture region where the flux-weighted centroids of the outflowing gas are located. The very high-resolution imaging capability of EVN at $18~\text{cm}$ resolves $< 10~\text{pc}$ scales. Figure~\ref{fig:EVN} shows one-sided pc-scale radio emission obtained using the EVN. We detect four compact radio knots (S0, S1, S2 and S3) over $3\sigma$ confidence, and in the brightest knot S0 we measure $S/N \sim 6$. 

We assume that the flux-weighted centroid of S0 is the radio core and align it with the astrometric BLR H$\beta$ center. The angular resolution of EVN is two orders of magnitude higher than MUSE WFM. To account for this difference, we use $3\sigma$ uncertainties of the astrometric locations of the flux-weighted H$\alpha$ and [\ion{O}{3}] centroids. In Figure~\ref{fig:EVN}, we show that S1 overlaps with the flux-weighted H$\alpha$ centroid, while S3 is in the vicinity of the flux-weighted [\ion{O}{3}] centroid. The radio knots are all directed toward the southwest, and thus appear to be aligned along one axis. The morphological parameters of the regions were estimated by fitting their flux profiles with a 2D Gaussian using the AIPS task JMFIT. They are summarized with their uncertainties in Table~\ref{table:radio_results}.

\subsubsection{Radio spectrum of the source}\label{subsubsec:radio-spectrum}

\begin{figure*}
    \centering
    \includegraphics[width=0.5\textwidth]{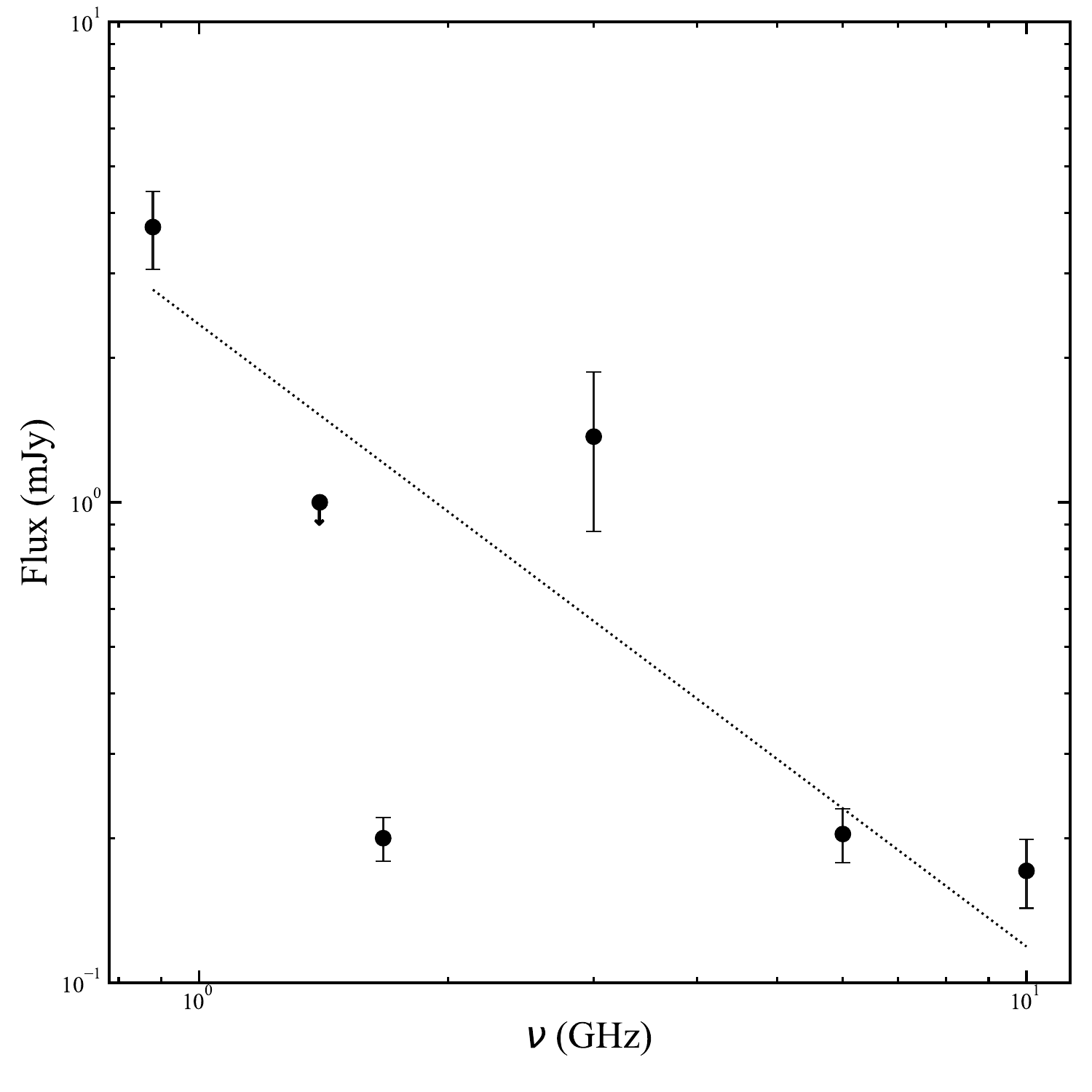}
    \caption{\textcolor{black}{Radio continuum spectrum for HE~0040$-$1105 between $880~\text{MHz}-10~\text{GHz}$, showing the variation of the integrated flux density values (mJy) with frequency. Note that we include only the upper limit for the VLA FIRST survey. The black dashed line indicates the best-fitted power-law model to the radio spectrum within the frequency range $880~\text{MHz}$ - $10~\text{GHz}$, yielding a spectral index of $\alpha_{\text{int, 0.88-10~GHz}} = -1.30 \pm 0.28$. We ignored the EVN observation during the fitting procedure as it could capture only a small fraction of the source's flux density. The radio spectrum is most consistent with a steep spectrum component.}}
    \label{fig:radio_spectra}
\end{figure*}

The $880~\text{MHz}-10~\text{GHz}$ radio spectrum of HE~0040$-$1105 (Figure~\ref{fig:radio_spectra}) is consistent with a straight spectrum, and a power-law fit yields a spectral index, \textcolor{black}{$\alpha_{\text{int, 0.88-10~GHz}} = -1.30 \pm 0.28$}. There is no evidence for either a peak or break in the spectrum over the observed frequency range. The integrated flux density ($S_{\text{int}}$) in the EVN observations is much lower than the expected $S_{\text{int}}$ assuming a power-law extrapolation from the lower resolution data. Such a discrepancy in $S_{\text{int}}$ suggests that the EVN is resolving out larger scale structure.

As mentioned earlier, the upper limit to its flux density at $1.4~\text{GHz}$ is $S_{\text{Thr, FIRST}} \sim 1~\text{mJy}$. However, the radio spectrum in Figure~\ref{fig:radio_spectra} \textcolor{black}{suggests an extrapolated flux density of $1.52 \pm 0.26~\text{mJy}$. This discrepancy is possibly due to  variability of the radio source since the FIRST observations were taken in 1995 and the VLA observations at $3$, $6$, and $10~\text{GHz}$ were taken in the last few years. Similar radio variability is seen in other radio quiet AGN \citep[e.g.,][]{Falcke2001, Nagar2002, Barvainis2005, Mundell2009}}.

\section{Discussion}\label{sec:discussion}

Combining the optical VLT/MUSE with \textit{HST} images has revealed multiple kinematic components in HE~0040$-$1105's ionized gas emission. Together with the radio-interferometric observations from VLA and EVN, we were able to resolve compact radio-emitting regions that are close to the galaxy nucleus from where an ionized gas outflow is launched. In this section, we combine spatial information from optical and radio observations to (i) discuss the origin of radio emission in Section~\ref{subsec:radio_outflow}, (ii) investigate the outflow launching mechanism in Section~\ref{subsec:outflow_origin}, (iii) explore whether the ionized gas outflow can escape the host galaxy in Section~\ref{subsubsec:outflow_escape}.

\subsection{Origin of the radio emission}\label{subsubsec:origin_of_radio_emission}

\subsubsection{AGN coronal emission}\label{subsubsec:coronal_emission}

If S0 is the AGN's radio core, then the AGN corona's synchrotron emission will contribute to the observed radio emission from S0. \citet{Laor2008} studied 71 radio-quiet and 16 radio-loud Palomar-Green AGN \citep{Green1986}. Radio luminosities ($L_{\text{R}}$) and X-ray luminosities ($L_{\text{X}}$) in the radio-quiet AGN follow $L_{\text{R}}/L_{\text{X}} \sim 10^{-5}$ \citep{Guedel1993}, which is usually seen in coronally active stars. The X-ray component is caused by thermal free-free emission from the hot $T \sim 10^{7}~\text{K}$ plasma, whereas the radio component is caused by nonthermal synchrotron emission. However, the coronal radio emission is produced on scales of a few hundred SMBH Schwarzschild radii \citep{Panessa2019,Wilkins2021}. The projected offset between S0 and the closest knot S1 is $\sim 15~\text{pc}$, which is a few orders of magnitude higher than where the radio emission from the corona would be produced. Therefore S1, S2, and S3 are not produced due to the coronal emission.

\subsubsection{Star forming processes}\label{subsec:radio_outflow}

To estimate whether HE~0040$-$1105's radio emission can be explained by processes that are related to star formation, we estimate the nuclear SFR and compare it with the value derived from the H$\alpha$ luminosity (Section~\ref{subsubsec:SFR_outflow}). Assuming that the entire radio emission originates from star formation-related processes, we utilize the VLA X-band observations to infer the central $1\arcsec$ SFR from where the emission originates. The integrated flux density at $10~\text{GHz}$ is $S_{\text{10~GHz}} \sim 171~\mu\text{Jy}$. Using the spectral index derived from the peak brightness, $\alpha \sim -0.6$, the integrated flux density at $1.4~\text{GHz}$ becomes $S_{\text{1.4 GHz}} \sim 556~\mu\text{Jy}$. The $1.4~\text{GHz}$ SFR ($\text{SFR}_{\text{1.4 GHz}}$) can be estimated following the calibration prescribed by \citet{Murphy2011}:
\begin{equation}
    \frac{\text{SFR}_{\text{1.4 GHz}}}{\text{M}_{\odot}~\text{yr}^{-1}} = 6.35 \times 10^{-29} \left( \frac{L_{\text{1.4 GHz}}}{\text{erg}~\text{s}^{-1}~\text{Hz}^{-1}} \right),
\end{equation}
where $L_{\text{1.4 GHz}}$ is the radio luminosity at $1.4~\text{GHz}$.

We estimate $\text{SFR}_{\text{1.4~GHz}} = 1.45~\text{M}_{\odot}~\text{yr}^{-1}$, which is more than two orders of magnitude higher than $\text{SFR}_{\text{O1}}$ derived in Section~\ref{subsubsec:SFR_outflow}. This suggests that supernovae and star-forming processes alone cannot explain the observed radio emission in the central $1\arcsec$ aperture. Therefore, at least one other process is required to explain HE~0040$-$1105's radio luminosity in the central $1\arcsec$ aperture region. The existence of supernovae clusters in the nuclear ($< 100~\text{pc}$) region cannot be easily ruled out.

The total $1.4~\text{GHz}$ radio luminosity of the knots S0, S1, S2, and S3 is $1.09 \times 10^{21}~\text{W Hz}^{-1}$. If we assume that the observed radio luminosity measured in these knots originates entirely from star formation-related processes, the predicted SFR of $0.69~\text{M}_{\odot}~\text{yr}^{-1}$ is still two orders of magnitude higher than $\text{SFR}_{\text{O1}}$. We conclude that S0-S3 are therefore not clusters of supernovae. The spatial alignment of the radio knots suggests that they are parts of a collimated radio-emitting structure.

\subsubsection{AGN wind vs. jet}\label{subsubsec:jet_driven_outflow}

\citet{Zakamska2014} proposed that the shocks from AGN-driven winds could give rise to non-thermal radio emissions. They suggested that the AGN wind could escape through the path of least resistance, and the radio emission produced by these winds would be diffuse. A recent study by \citet{Liu2022} using the Very Large Baseline Array (VLBA) has suggested that radio emission produced by AGN winds may not contain any compact structures. On the other hand, we detect compact radio knots on pc-scales in our EVN observations. If the findings of \citet{Liu2022} are true, the pc-scale knots detected in our EVN observations are likely to be a jet. However, these radio structures lack a traditional linear jet-like morphology as seen \citep[e.g.,][]{Huchra1992,Brunthaler2000,Thean2001}. Another possibility is that the jets and the winds could operate at the same time. \citet{Agudo2015} suggested that if the mass loading of the disk-driven wind cannot produce an ultra-relativistic flow, both jets and winds could operate simultaneously.

\subsection{Driver of the outflow}\label{subsec:outflow_origin}

\subsubsection{Constraints from the outflow energetics}\label{subsection:energetics_comparison}

The radio properties of the observed emission cannot explain its connection to the outflow. A one-to-one connection could only be established if the momentum and kinetic energy injection by the wind or jet is greater than those of the outflowing ionized gas. The momentum injection due to AGN radiation can be calculated as $\dot{P}_{\text{AGN}} \sim L_{\text{bol}}/c \sim 4.6 \times 10^{33}~\text{dyne}$. In Figure~\ref{fig:energetics}, $\dot{P}_{\text{ion, cone}}/\dot{P}_{\text{AGN}} < 1$ for all values of $v_{\text{out}}$, while $\dot{P}_{\text{ion, shell}} < \dot{P}_{\text{AGN}}$ only for $v_{\text{out}} = v_{\text{broad}}$ when $\Delta R > 180~\text{pc}$. This momentum boost indicates that the outflow is likely to be energy-conserving \citep{Faucher-Giuere12}. $L_{\text{bol}}$ can easily produce the observed injection rate of kinetic energy from the outflow. We find that $\sim 20\%$ of the optical bolometric luminosity of the AGN can produce the maximum predicted kinetic energy injection rate $\dot{E}_{\text{ion, max}} \sim 3 \times 10^{43}~\text{erg~s}^{-1}$. Therefore, the AGN radiation pressure can drive these outflows.

However, the existence of the compact knot-like feature in our EVN observation is more likely to be produced by a jet. Assuming that the entire observed VLA radio emission is due to a jet, the estimated $1.4~\text{GHz}$ radio luminosity is $\log{(L_{\text{1.4~GHz}}/\text{W Hz}^{-1})} \sim 21.37$. Using the scaling relations of radio power versus jet power given by \citet{Cavagnolo2010}, we estimate that the associated jet power ($P_{\text{jet}}$) is $\sim 1.3 \times 10^{42}~\text{erg~s}^{-1}$, with an upper limit of $5.8 \times 10^{42}~\text{erg~s}^{-1}$. $P_{\text{jet}}$ is about an order of magnitude higher than $\dot{E}_{\text{ion,~cone}}$ and $\dot{E}_{\text{ion, shell}}$ except when $v_{\text{out}} = v_{\text{max}}$.

\begin{figure*}
    \centering
    \includegraphics[width=\textwidth]{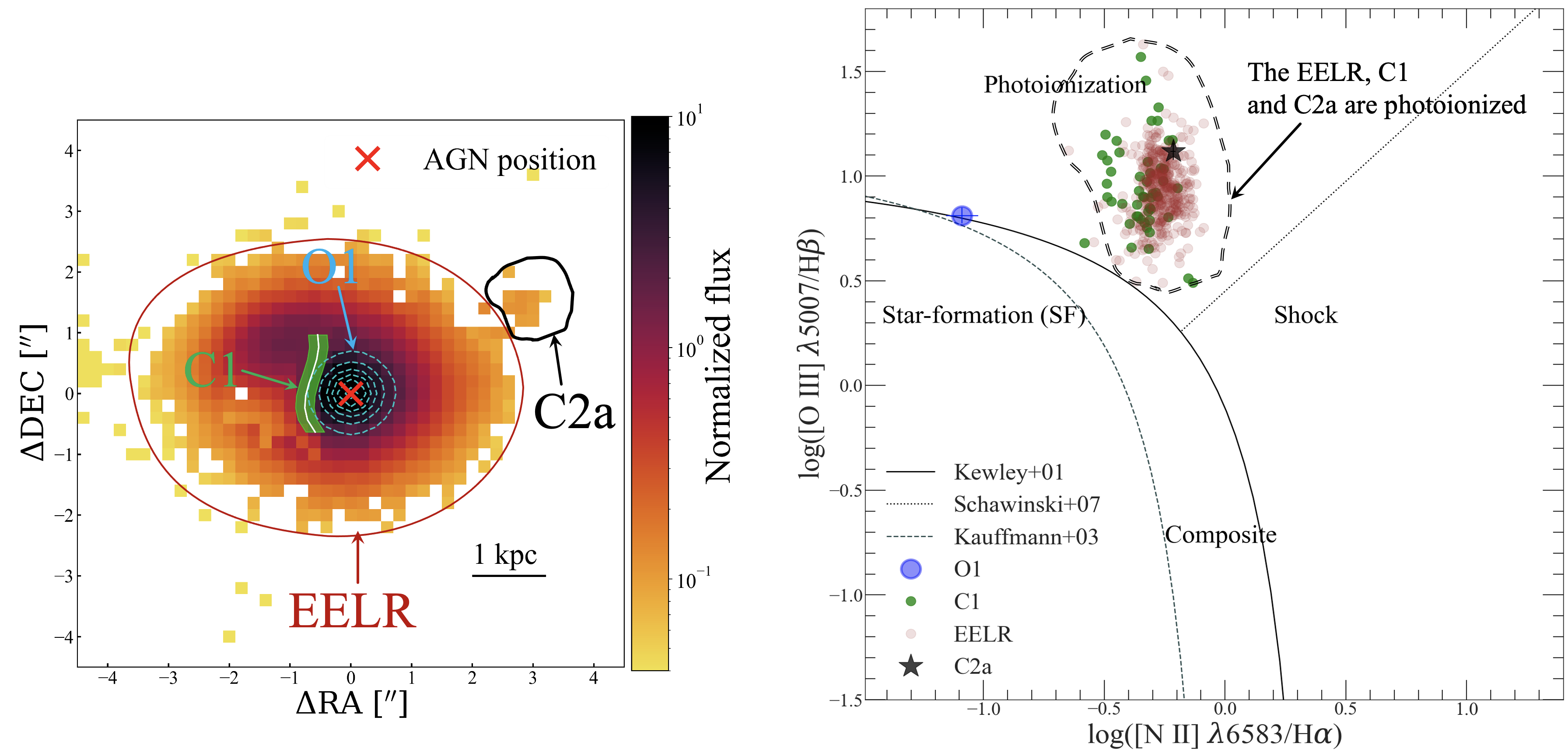}
    \caption{BPT diagram showing the [\ion{O}{3}]~$\lambda 5007$/H$\beta$ versus [\ion{N}{2}]~$\lambda 6583$/H$\alpha$ ratios for different regions across the entire ionized nebula. \textit{Left panel}: [\ion{O}{3}] flux map with the extraction of different regions shown by curved lines and arrows. The red cross denotes the AGN location. \textit{Right panel}: The black dotted line represents the upper limit of ionization due to pure star formation. The region between the black dotted line and the black solid line represents ionization due to both star formation and AGN. The area within the black and gray solid lines represents ionization due to shock. Above the black and gray solid lines, the hard ionization photo-ionizing field from the AGN dominates the line ratios. The blue cross and the black stars describe the regions O1 and C2a, respectively, and are plotted with error bars with similar colors. The green and brown dots represent the feature C1 and the EELR and are not plotted with error bars because they are spatially resolved. O1 resides between the composite and AGN photoionized region within uncertainties. The EELR, C1, and C2a are photoionized by the AGN.}
    \label{fig:BPT}
\end{figure*}

\subsubsection{Constraints from the ionization mechanism}\label{subsubsec:ionization_outflow}

We estimate $\log{([\text{\ion{O}{3}}]/\text{H}\beta)} = 0.81 \pm 0.04$ and $\log{([\text{\ion{N}{2}}]/\text{H}\alpha)} = -1.09 \pm 0.07$, which places O1 between the composite and AGN-photoionized regions in the BPT diagram \citep[][see Figure~\ref{fig:BPT}]{Baldwin1981}. As we have discussed in Section~\ref{subsubsec:coronal_emission}, the energy and momentum contribution from star-forming-related processes is not sufficient to power the ionized gas outflow. To test whether the AGN radiation can power the O1 outflow we estimate the ionization parameter $U$ which represents the ratio of ionizing photon flux to the electron density of the ambient medium, and is defined as
\begin{equation}
    U = \frac{Q}{4\pi r^{2} n_{\text{H}} c},
\end{equation}
where, $L_{\text{ion, O1}}$ is the luminosity of the ionizing source, $\nu$ is the frequency of the ionizing radiation, $h$ the Planck constant, $r$ the distance from the ionizing source, and $n_{\text{H}} \sim 0.85 n_{\text{e}}$ the hydrogen number density per unit area \citep{Crenshaw2015}. The ionization parameter $U$ can be estimated from the emission line ratios \citep{Baron2019DiscoveringLocation}
\begin{align*}
    \log{U} = & + 0.191 \left[ \log{\left( \frac{\text{[\ion{O}{3}]}}{\text{H}\beta} \right)} \right] + 0.778 \left[ \log{ \left( \frac{\text{[\ion{O}{3}]}}{\text{H}\beta} \right)} \right]^{2} \\
    & - 0.251 \left[ \log{ \left( \frac{\text{[\ion{N}{2}]}}{\text{H}\alpha} \right)} \right] + 0.342 \left[ \log{ \left( \frac{\text{[\ion{N}{2}]}}{\text{H}\alpha} \right)} \right]^{2} \\
    & + 3.766.
\end{align*}

For the ionized gas outflow O1, we estimate $U$ and the associated uncertainty with a Monte-Carlo approach, yielding $\log{U} = -2.42 \pm 0.08$.

As we use the $5100~\text{\AA}$ continuum luminosity to estimate the optical bolometric luminosity of the AGN, the corresponding frequency at that wavelength is  $\nu = c/(5100~\text{\AA}) = 5.77 \times 10^{14}~\text{Hz}$. Using $Q= L_{\text{ion}} /h\nu$, we  estimate the ionizing luminosity as $L_{\text{ion, O1}} = (1.05-1.1) \times 10^{43}~\text{erg s}^{-1}$. This value is an order of magnitude lower than $L_{\text{bol}}$, indicating that the bolometric luminosity of the AGN may not be able to photoionize the gas clouds, i.e., is not able to produce the emission line ratios observed in O1.

A scenario in which the AGN radiation may still be the powering mechanism is provided by optically thick gas clouds blocking a significant fraction of AGN radiation. This may lead to an AGN luminosity that is higher than the estimated value from the emission line ratios $L_{\text{ion, O1}}$. We estimate the column density in O1 ($N_{\text{H, O1}}$) using the $N_{\text{H}} - A_{\text{V}}$ scaling relation from \citet{Guver2009}, as follows:
\begin{equation}
    N_{\text{H, O1}} = (2.21 \pm 0.09) \times 10^{21} A_{\text{V, O1}}~\text{cm}^{-2},
\end{equation}
yielding $N_{\text{H, O1}} = (6.0 \pm 1.5) \times 10^{21}~\text{cm}^{-2}$, i.e., an optically thin outflow O1. The clouds only become optically thick when they reach a hydrogen column density value $>10^{23}~\text{cm}^{-2}$ \citep{Jaffarian2020}. Therefore, the optically thin clouds may not be able to block $> 90\%$ of the incoming ionizing luminosity from the AGN and produce the observed low value of $L_{\text{ion, O1}}$. We note, however, that there have been other mechanisms proposed that could shield radiation from the AGN, but not be reflected in our estimate for $N_{\text{H}}$. They include self-shielding from ionized gas \citep{Zubovas2017} and screening due to the shadow of the host disk \citep{Husemann2019}.

\subsubsection{O1 outflow powering mechanism}

We now explore the possibility of weak radio jets being the ionizing source of the O1 outflow. Although the upper limit of $P_{\text{jet}}$ is a factor of two lower than $L_{\text{ion, O1}}$ at $r = 480~\text{pc}$ (Section~\ref{subsubsec:jet_driven_outflow}), $P_{\text{jet}}$ agrees well with the possible attainable values for $L_{\text{ion, O1}}$ estimated at the location of the ionized gas outflow at $r = 150~\text{pc}$ , which is in the vicinity of the radio source S1. In \citet{Singha2021a}, we found that $(99 \pm 1)\%$ of the observed outflowing [\ion{O}{3}] emission is located at the flux-weighted centroid. For an AGN wind-driven outflow, as the wind propagates along every possible direction, the bulk of the [\ion{O}{3}] emission is unexpected to be located at one particular region. Together with the close proximity of the entirety of the ionized gas outflow emission to the collimated radio jet-like structure, where the comparability of the ionizing luminosity of the outflowing gas is comparable to the jet power, we propose that weak radio jets could drive this ionized outflow at its location. However, AGN radiation is necessary to carry the gas to larger distances, which is necessary to explain the high ionizing luminosity $L_{\text{ion, O1}}$ on $500~\text{pc}$ scales.

We propose that the radio jets could transfer their mechanical energy to the ambient gas, and therefore perturb and ionize the clouds in their vicinity. As they move outward with the photons from the ionization cone, they gradually become photoionized, which is in line with the two-stage acceleration mechanism proposed by \citet{Hopkins2010}. Our analysis of the ionized gas outflow is limited by both sensitivity and finite spatial resolution, which also limits our understanding of the complex interaction between radio emission, AGN radiation field, and the host galaxy ISM in HE~0040$-$1105.

\citet{Irina2021} found that the large ionized gas nebula EELR is largely photoionized. However, their analysis did not involve the resolved analysis of the ionized gas outflow O1 and the clouds C1 and C2a. In the BPT diagram shown in Figure~\ref{fig:BPT} it becomes evident that each of the clouds C1 and C2a is photoionized.

\subsection{Outflow escaping the host galaxy gravitational potential}\label{subsubsec:outflow_escape}\label{subsecion:outflow_escaping_potential}

While numerous studies have shown that AGN-driven outflows are able to deprive galaxies of their gas component \citep[e.g.,][]{Benson2003,Bower2006,Moll2007,Schindler2008,McCarthy2010,Gaspari2011,Crenshaw2012,McNamara2012,Schaye2015,Gaspari:2018}, there is a significant fraction of the outflowing material that falls back and may eventually be accreted by the AGN \citep{Oppenheimer2010,Diniz2015,Gaspari:2013,Muratov2015,Gaspari:2017,Wittor:2020}. To test whether HE~0040$-$1105's outflowing ionized gas can escape the host galaxy's gravitational potential, we assume an inclination $i = 40\arcdeg$ and estimate its escape velocity ($v_{\text{esc, ion}}$) of the ionized gas from the host, and assume an inclination $i = 40\arcdeg$ throughout the entire analysis. We follow the prescription of \citet{Rupke2002} which uses the gravitational model of an isothermal sphere to estimate $v_{\text{esc, ion}}$ at a distance $r$ from the AGN nucleus as
\begin{equation}
    v_{\text{esc, ion}}(r, r_{\text{max}}) = \sqrt{2} v_{\text{rot}} \left[ 1 + \ln{\left( r_{\text{max}}/r \right)} \right]^{1/2},
\end{equation}
where $v_{\text{rot}}$ is the rotational velocity of the host galaxy. Following \citet{Villar-Martin2018}, we estimate $v_{\text{rot}} = \sqrt{2} \sigma_{\text{*}} \sim 181~\text{km s}^{-1}$.

In the above, $r$ represents the distance from the AGN nucleus and $r_{\text{max}}$ is the maximum radius of the dark matter halo, which is observationally unconstrained. We estimate $r_{\text{max}}$ from the scaling relation \citep{Huang2017} between the virial radius of the dark matter halo and the deprojected host galaxy effective radius $R_{\text{eff}} = 9.6~\text{kpc}$ \citep{Husemann2021} as $r_{\text{max}} \sim 420~\text{kpc}$. Our estimate for $r_{\text{max}}$ should be regarded as an upper limit since studies of outflows in quasar host galaxies often assume $r_{\text{max}} = 100~\text{kpc}$ \citep[e.g.,][]{Greene2011,Herrera-Camus20192,Villar-Martin2018}. For the escape velocity at the outer boundary of O1 where $r = 480~\text{pc}$ and $r_{\text{max}} = 420~\text{kpc}$ we estimate $v_{\text{esc}} \sim 710~\text{km s}^{-1}$, which is higher than the maximum velocity of the outflow $v_{\text{max}}$ (see Table~\ref{tab:gas_cloud_properties}). For $r_{\text{max}} = 420~\text{kpc}$, the escape velocity at $r= 480~\text{pc}$ is $v_{\text{esc}} \sim 640~\text{km s}^{-1}$, implying that the O1 outflow is barely able to escape the central region. For the other outflow velocity values (Table~\ref{tab:energetics}), the ionized gas is unable to escape from the gravitational potential of the host galaxy on $< 500~\text{pc}$ scales from the nucleus, and hence cannot reach kpc scales. This is also consistent with the k-plot diagnostic shown in Figure~\ref{fig:k-plot}. As discussed in Section~\ref{subsubsec:CCA}, such gas is expected to condense in-situ and soon rain back toward the central SMBH via the CCA mechanism. This creates a self-regulated feeding and feedback AGN loop, which is at the core of currently consistent theoretical models \citep[][for a review]{Gaspari:2020}.

\section{Conclusions}\label{sec:conclusion}

In this work, we have presented a multi-wavelength analysis of the ionized gas outflow in the nearby, radio-quiet AGN HE~0040$-$1105. The observations in the optical, UV and radio have revealed multiple features that extend from galaxy scales down to $< 10~\text{pc}$ from the AGN. Our key findings are summarized as follows:

\begin{itemize}
    \item The ionized gas outflow in H$\alpha$ is spatially unresolved by MUSE (Section~\ref{subsubsec:identifying_Ha_outflow}) and confined within the central $500~\text{pc}$ from the nucleus (Section~\ref{subsubsec:outflow_size}). For this region, we estimate an upper limit for the SFR (Section~\ref{subsubsec:SFR_outflow}), which is exceeded by the mass outflow rate by two orders of magnitude.
    
    \item The kpc-scale ionized nebula comprises four kinematically distinct regions: (i) a central blue wing/ionized gas outflow O1, (ii) a receding ionized gas shell C1, (iii) the EELR that is local to the galaxy, and (iv) a blue-shifted knot C2a on the northwest side of the nucleus (Section~\ref{subsec:resolved_host_galaxy_emission}). Although the kinematics of the O1 outflow is non-gravitational, the ionized gas motion on large scales is dominated by the bulge potential of the galaxy. The kinematic misalignment between the stars and H$\alpha$ ionized gas, the quiescent EELR kinematics and the detection of the clumpy gas in the UV is consistent with the idea that HE~0040$-$1105 is a late state merger, which can further enhance CCA condensation.
    
    \item The flux-weighted centroids of the outflowing gas in H$\alpha$ and [\ion{O}{3}] coincides with radio emission detected in the observations acquired with the VLA and EVN (Section~\ref{subsec:radio_emission}). Our findings suggest that the observed spatial alignment of the pc-scale radio knots is consistent with a weak radio jet morphology, as opposed to the diffuse radio emission predicted for AGN winds (Section~\ref{subsubsec:jet_driven_outflow}). 

    \item The radio spectrum of HE~0040$-$1105 within the frequency range of $880~\text{MHz}$-$10~\text{GHz}$ is consistent with a steep spectrum \textcolor{black}{($\alpha = -1.30 \pm 0.28; S_{\nu} \propto \nu^{\alpha}$)}. Additionally, the source demonstrates radio continuum variability on a $\sim 20~\text{yr}$ timescale (Section~\ref{subsubsec:radio-spectrum}).
    
    \item The majority ($\sim 99\%$) of the [\ion{O}{3}] emission is located at the location of the flux weighted [\ion{O}{3}] centroid, where the estimated ionizing luminosity of the O1 outflow is similar to the mechanical power of the radio jets. However, at the outer boundary of O1, the ionizing luminosity is a factor of two higher than the jet power but an order of magnitude lower than the AGN bolometric luminosity. The most conclusive scenario is that the radio jets accelerate and ionize the ambient medium through the dissipation of their mechanical energy on $100~\text{pc}$ scales, while on galaxy scales ($\sim 500~\text{pc}$) the gas clouds interact with the photons from the AGN radiation field and are therefore photoionized (Section~\ref{subsubsec:ionization_outflow}).
    
    \item The velocity of the O1 outflow is too low to escape the host galaxy gravitational potential from the central $500~\text{pc}$, in which case the ionized clouds may rain back onto the SMBH via CCA (Section~\ref{subsecion:outflow_escaping_potential}). Indeed, as tested by the k-plot diagnostic, most of the ionized clouds in HE~0040$-$1105 are expected to be prone to multi-phase turbulent condensation in situ (Figure~\ref{fig:k-plot}).
\end{itemize}

Our results stress the complexity of the outflow-ISM interaction on different spatial scales, as well as the challenge to constrain the launching mechanism of nuclear-ionized gas outflows in radio-quiet AGN. Although VLBI allowed us to tenuously resolve a jet-like morphology in the faint radio emission from HE~0040$-$1105, it can not be directly identified with the galaxy-scale processes in the traced ionized gas. To distinguish between the AGN-wind versus jet launching mechanism, further multi-wavelength observations are required to spatially resolve the kinematic features. Only resolving the multi-phase ISM will help to understand the complex interaction between AGN and host galaxy on different spatial scales.

\begin{acknowledgments}
We thank the anonymous referee for providing useful comments which have improved the overall quality of the paper. The work of MS was supported in part by the University of Manitoba Graduate Enhancement of Tri-Council Stipends (GETS) program and by NASA under award number 80GSFC21M0002.
The work of CPO and SAB was supported by a grant from the Natural Sciences and Engineering Research Council (NSERC) of Canada. MPT acknowledges financial support from the State Agency for Research of the Spanish MCIU through the ``centre of Excellence Severo Ochoa'' award to the Instituto de Astrofísica de Andalucía (SEV-2017-0709) and through grant PID2020-117404GB-C21 (MCI/AEI/FEDER, UE). MG acknowledges the partial support by NASA Chandra GO9-20114X and HST GO-15890.020/023-A, and the {\it BlackHoleWeather} program. The Science and Technology Facilities Council is acknowledged by JN for support through the Consolidated Grant Cosmology and Astrophysics at Portsmouth, ST/S000550/1.

The European VLBI Network is a joint facility of independent European, African, Asian, and North American radio astronomy institutes. Scientific results from data presented in this publication are derived from the following EVN project code: EP119 (PI: P{\'e}rez-Torres). e-VLBI research infrastructure in Europe is supported by the European Union’s Seventh Framework Programme (FP7/2007-2013) under grant agreement number RI-261525 NEXPReS. The National Radio Astronomy Observatory is a facility of the National Science Foundation operated under cooperative agreement by Associated Universities, Inc.
\end{acknowledgments}

\facilities{EVN, VLA, VLT:Yepun (MUSE)}

\software{Astropy \citep{astropy2013,astropy2018}, CASA \citep{McMullin2007}, AIPS \citep{Wells1985}, Matplotlib \citep{Hunter2007}, NumPy \citep{VanDerWalt2011}, SciPy \citep{Virtanen2020}, VorBin \citep{Cappellari2003}}

\bibliography{ms.bib}

\pagebreak

\appendix

\section{Fitting multi-component emission line spectra in the H$\alpha$ window}\label{appendix:Ha_NII}

\begin{figure*}
    \centering
    \includegraphics[width=0.45\textwidth]{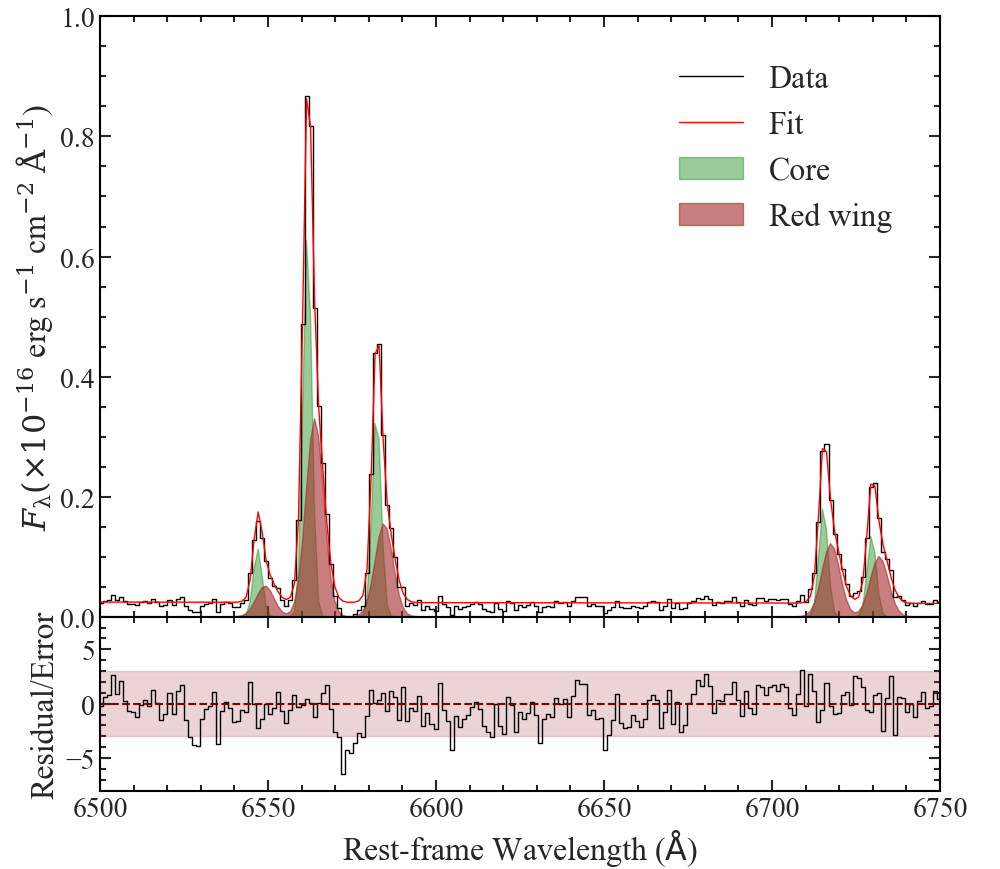}
    \includegraphics[width=0.45\textwidth]{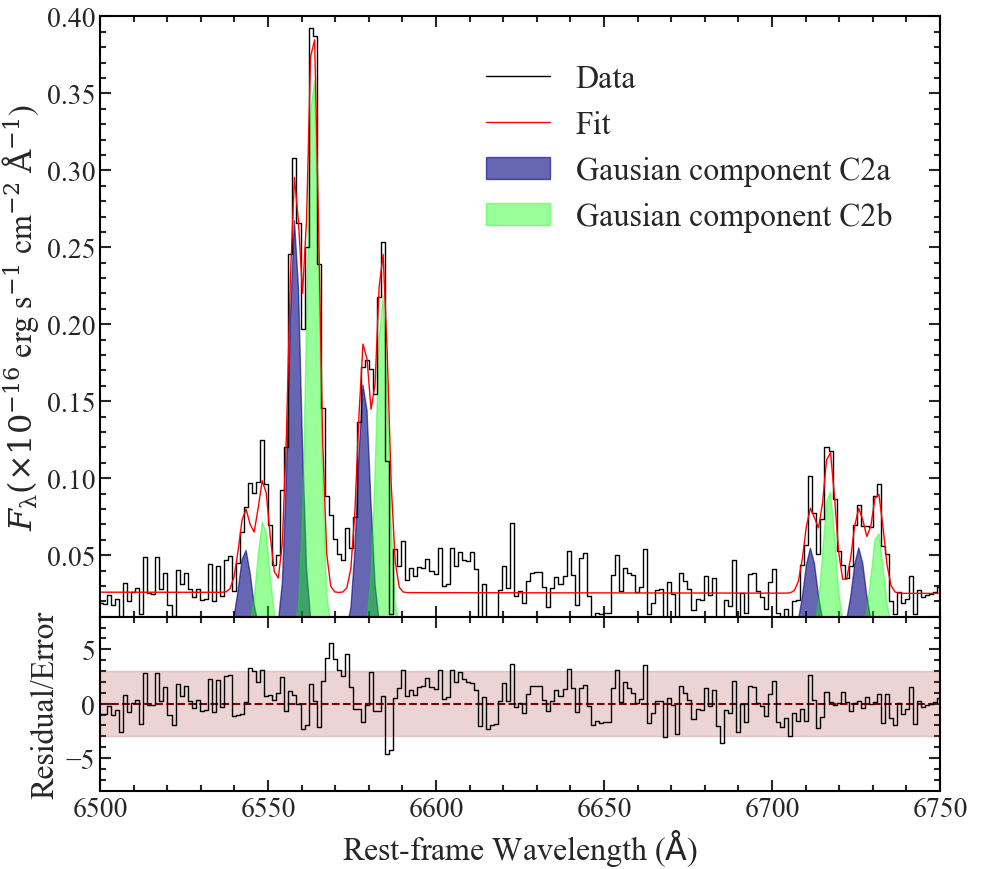}
    \caption{Similar to the left panel of Figure~\ref{fig:C1}, the left panel here shows the $3 \times 3$ spaxel aperture spectra extracted from C1 and their two-component modelling. The figure organization and symbols are similar to Figure~\ref{fig:C1} (left panel). The right panel shows the $9 \times 9$ spaxel aperture spectra extracted at the location of C2 together with the best-fit two-component model and the emission line spectra of the individual components. The figure organization and symbols are similar to the upper left panel of Figure~\ref{fig:knot_maps}.}
    \label{fig:appendix_Ha_spectra}
\end{figure*}

In Section~\ref{subsubsec:C1} and Section~\ref{subsubsec:C2}, we have spectroscopically deblended the red-wing components of both features C1 and C2 in both [\ion{O}{3}] and H$\alpha$ windows. While we have only shown the spectral fit for the H$\beta$ + [\ion{O}{3}] emission lines, the BPT diagnostic also requires the emission line fluxes of H$\alpha$ and [\ion{N}{2}]. Since the lines consist of multiple components, we also show their spectra and the best-fit models here in Figure~\ref{fig:appendix_Ha_spectra}. Similar to the modelling of the H$\beta$ and [\ion{O}{3}] emission lines shown in Figure~\ref{fig:C1}, we extract the same $3 \times 3$ spaxel aperture spectrum around the H$\alpha$ + [\ion{N}{2}] + [\ion{S}{2}] emission line complex and fit it with a two-component model. The second component is only employed if the AIC criterion described in Section~\ref{subsubsec:identifying_Ha_outflow} is fulfilled.

We model the $9 \times 9$ spaxel aperture spectrum around C2 (see Figure~\ref{fig:knot_maps}) in the H$\alpha$ window in a similar fashion as for C1. 

In both cases, C1 and C2, two components are required to reproduce the emission line shape. In the C1 aperture spectrum, the core component corresponds to the EELR local to the galaxy, whereas the red-wing belongs to the kinematically distinct feature C1 (denoted by the brown Gaussian components). For the C2 aperture spectrum, the blue and the green Gaussian components correspond to C2a and C2b respectively. Combining the best-fit results with what we have retrieved in the H$\beta$ + [\ion{O}{3}] window, the BPT characterization of C1, C2a, and C2b are shown in Figure~\ref{fig:BPT}.

\section{Constraining the Morphology of the ionized gas cloud C1}\label{appendix:deconvolution}

\begin{figure}[ht]
    \centering
    \includegraphics[width=0.9\textwidth]{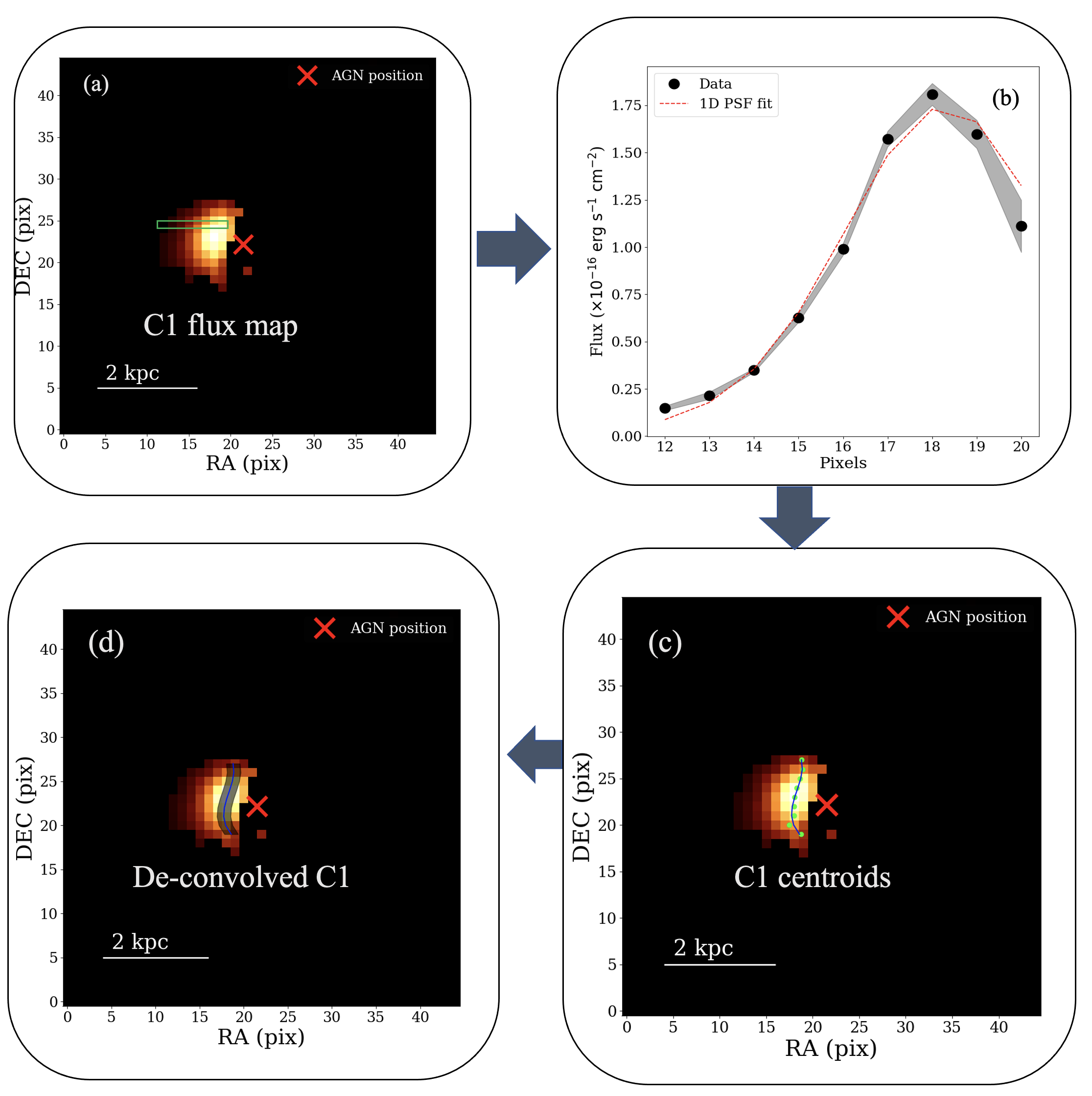}
    \caption{A flowchart showing the method by which we constrain the location and morphology of the ionized gas cloud C1. Panel (a) shows the flux map retrieved from the two-component model fitted to the original MUSE data cube. In panel (b) we show an example slice of the flux profile at fixed DEC, together with the model of the PSF. This slice has been taken from the slice at the DEC with the brightest C1 emission, where the 1D PSF fit follows the 1D flux profile closely. We map the location of the one-dimensional surface profiles in panel (c), which line up to a shell that extends in the north-south direction. The blue line represents the centroids of C1. Panel (d) shows the deconvolved image of C1, where the width of the underlying structure is achieved by fitting with the flux profile at a fixed DEC with the brightest C1 emission with a 1D PSF convolved with a 1D Gaussian. The standard deviation of the Gaussian provides us with the width of the structure (shaded in gray).}
    \label{fig:appendix_flowchart}
\end{figure}

In Section~\ref{subsubsec:stellar_velocity_field_EELR} we have identified a spatially resolved, redshifted ionized gas cloud C1 that is located east of the AGN. Here we describe how we constrain the underlying morphology. As a first step, we map the surface brightness profile of the emission line component which is shown in the left panel of Figure~\ref{fig:appendix_flowchart}. The red wing component's signal is blended with the bright blue wing component. The AIC criterion described in Section~\ref{subsubsec:identifying_Ha_outflow} therefore introduces an artificial cut-off close to the nucleus which is not physical. Since the peak of the surface brightness profile is clearly visible and offset from the AGN position, we exclusively concentrate on the eastward extension of the profile for the following analysis. Since C1 appears to be elongated in the north-south direction, we first extract the 1D flux of the source at fixed DEC. As a next step, we estimate the width of the source by fitting the profile with a one-dimensional Gaussian profile, for every slice extracted. In order to account for beam smearing, we convolve the Gaussian profile with the PSF before minimizing the $\chi^2$ residuals, which provides us with both the location and the intrinsic extent of the feature. The bottom right panel of Figure~\ref{fig:appendix_flowchart} shows that the location of C1's maximum luminosity is almost constant, which we interpret as a shell-like morphology. The shell is located $600~\text{pc}$ distance away from the nucleus. Its median width in the east-west direction is $255~\text{pc}$.

\end{document}